\documentclass[
  aps,
  pra,
  twocolumn,
  superscriptaddress,
  nolongbibliography,
  10pt,
]{revtex4-2}

\usepackage{graphics}
\usepackage{graphicx}
\usepackage{bbm,bm}
\usepackage{amsfonts}
\usepackage{amsmath}
\usepackage[dvipsnames]{xcolor}
\usepackage{nicefrac} 
\usepackage{siunitx}
\usepackage{braket}
\usepackage{bm}
\usepackage{tabularx}
\usepackage{array}

\ifx\pdftexversion\undefined
\usepackage[
  dvips
]{hyperref}
\else
\usepackage[
]{hyperref}
\fi
\hypersetup{
	hidelinks,
	colorlinks=true,
	linkcolor=black,
	citecolor=MidnightBlue,
	urlcolor=MidnightBlue,
}
\makeatletter
\def\maketitle{
\@author@finish
\title@column\titleblock@produce
\suppressfloats[t]}
\makeatother

\newcommand{\Ef}{E_\mathrm{f}}
\newcommand{\Ec}{E_\mathrm{c}}
\newcommand{\VR}{V_\mathrm{R}}
\newcommand{\trf}{t_\mathrm{rf}}

\newif\ifsupplement
\supplementfalse

\begin{document}
\preprint{}

\title{
Direct Observation of the Three-Dimensional Anderson Transition with Ultracold Atoms in a Disordered Potential}

\newcommand{\LCF}{Universit\'e Paris-Saclay, Institut d'Optique Graduate School, CNRS, Laboratoire Charles Fabry, 91127, Palaiseau, France}
\newcommand{\LKB}{Laboratoire Kastler Brossel, Sorbonne Universit\'e, CNRS, ENS-PSL Research University, Coll\`ege de France, 4 Place Jussieu, 75005 Paris, France }
\newcommand{\STE}{Universit\'e Jean Monnet Saint-Etienne, CNRS, Institut d'Optique Graduate School,
Laboratoire Hubert Curien, UMR 5516, Saint-Etienne F-42023, France }
\newcommand{\equalContribution}{These authors contributed equally to this work.}

\author{Xudong Yu}
\thanks{\equalContribution}
\affiliation{\LCF}

\author{Ke Xie}
\thanks{\equalContribution}
\affiliation{\LCF}

\author{Hoa Mai Quach}
\affiliation{\LCF}

\author{Yukun Guo}
\affiliation{\LCF}

\author{Myneni Niranjan}
\affiliation{\LCF}

\author{Sacha Barr\'e}
\affiliation{\LCF}

\author{Jean-Philippe Banon}
\affiliation{\LCF} \affiliation{\LKB}\affiliation{\STE}

\author{Alain Aspect}
\affiliation{\LCF}

\author{Nicolas Cherroret}
\affiliation{\LKB}

\author{Vincent Josse}
\email[Electronic address: ]{vincent.josse@institutoptique.fr}%
\affiliation{\LCF}

\begin{abstract}
Anderson localization of particles -- the complete halt of wave transport through multiple scattering and phase coherence -- is a paradigmatic manifestation of quantum interference in disordered media. In three dimensions, the scaling theory predicts a quantum phase transition at a critical energy, the mobility edge, separating localized from diffusive states and underpinning metal-insulator transitions in electronic systems. Despite decades of experimental efforts, a direct observation of this emblematic transition for matter waves has remained elusive. Previous attempts with ultracold atoms were hindered by strong and uncontrolled energy broadening, resulting in indirect, sometimes inaccurate, and model-dependent estimates of the mobility edge. Here we implement a novel energy-resolved scheme to prepare atomic matter waves with a narrow energy distribution and track their expansion dynamics over long timescales. This allows for a direct observation of the three-dimensional Anderson transition in a laser-speckle disordered potential, and for a precise measurement of the mobility edge that is independent of any underlying theoretical modeling. Our measurements show excellent agreement with state-of-the-art numerical predictions over a wide range of disorder strengths, resolving long-standing discrepancies between prior experiments and theory. Beyond the three-dimensional Anderson transition, our approach opens new avenues for quantitative investigations of quantum critical phenomena in spatially disordered systems, including the roles of dimensionality, symmetry class, and interactions.
\end{abstract}

\maketitle

In 1958, P.~W. Anderson showed that sufficiently strong disorder in a lattice can localize quantum particles, preventing their diffusion over large scales \cite{anderson1958}. This counter-intuitive effect -- now known as Anderson localization -- was later understood as a generic interference phenomenon for waves in disordered media~\cite{lagendijk2009fifty}. Since then, it has been explored in a wide variety of systems, ranging from electron transport in solids~\cite{Lee1985} to classical waves such as light, sound, and microwaves in random structures~\cite{Schwartz2007, segev2013,chabanov2000,acoustic-weaver1990anderson, Hu2008,yamilov2023}. 
In parallel, ultracold atomic gases have emerged as a powerful and versatile platform for studying disordered quantum systems, combining their general controllability~\cite{Bloch_Dalibard2012} with the ability to engineer disorder with well-defined statistical properties~\cite{clement2006experimental,sanchez2010disordered}. 
These assets have established these systems as key quantum simulators of localization phenomena~\cite{billy2008direct,roati2008anderson,kondov2011three,Fred2012,G.Semeghini2015,White2019_2DAL}, extending well beyond the original Anderson framework to include interacting~\cite{Bloch2015MBL, choi2016, Abanin2019}, dissipative~\cite{yusipov2017, huang2020, guo2025, yang2025}, and gauge-field-dressed systems~\cite{Hainaut2018, Gadway2018,Meier2019topologicalInsulator,Nakajima2021}, as well as novel theoretical perspectives such as the localization-landscape approach~\cite{filoche2012,filoche2024}.


A crucial feature of the Anderson problem is the role played by dimensionality. In one and two dimensions, arbitrarily weak disorder localizes all eigenstates, whereas in three dimensions a genuine quantum phase transition appears \cite{abrahams1979}. Experimental access to this transition has proven particularly challenging: conductivity measurements in electronic conductors are often hindered by phonon coupling, band hybridization or electron-electron interactions~\cite{Rosenbaum1980, Paalanen1982,Stupp93,Itoh2004,Carnio2019,ying2016anderson,Sauty2022}. For classical-wave systems, the phenomenon long remained controversial, with only one truly convincing observation reported to date~\cite{Hu2008}.
Cold-atom realizations of the quantum kicked rotor have provided an alternative route to probing Anderson-type criticality in momentum space, using temporal disorder and synthetic dimensions to emulate an effective three-dimensional (3D) dynamics~\cite{casati1989, chabe2008, lopez2012, madani2025}. Although these systems can be mapped onto the 3D Anderson model, they fundamentally differ from real-space Anderson localization, in which the transition occurs at a critical energy -- the mobility edge -- separating localized (insulating) and diffusive (conducting) states~\cite{abrahams1979}. 
Mobility edges in disordered systems have recently attracted considerable interest, even beyond the 3D Anderson transition, as they arise in different symmetry classes~\cite{evers2008,Slevin2014}, in one-dimensional quasiperiodic models~\cite{Biddle2010,Luschen2018,An2021MobilityEdge,Wang2022,Xu2024Reentrant}, and have been predicted to separate thermal from non-ergodic phases in the many-body counterpart of Anderson localization~\cite{Alet2018}. Direct experimental observation of mobility edges nevertheless remains highly challenging, typically requiring energy-resolved probes or similarly selective techniques~\cite{Roushan2017}.
For ultracold atoms in 3D spatial disorder, pioneering experiments have reported evidence of Anderson localization~\cite{kondov2011three,Fred2012,G.Semeghini2015}, but they all suffered from a major limitation: a strong broadening of the atomic energy distribution, which spanned across the Anderson transition. 
As a result, these early studies were limited to indirect estimates of the mobility edge, based on underlying models. This led to substantial and model-dependent discrepancies with theoretical predictions~\cite{Kuhn_2007,Skipetrov2008,yedjour2010diffusion,shapiro2012cold,Piraud_2012,Piraud2014,muller2014comment} and, ultimately, with state-of-the art numerical predictions~\cite{delande2014mobility,Ec_Delande2017}.

Here, we report the direct observation of the 3D Anderson transition, together with a precise determination of the mobility edge. This work relies on a new state-dependent loading scheme that allows us to prepare matter waves at a well-defined energy inside the disordered potential \cite{lecoutre2022bichromatic}, thereby overcoming limitations of previous attempts. By combining long-time spatial dynamics with this precise energy selectivity, we achieve a purely experimental determination of the critical point of the Anderson transition 3D laser speckle disorder, without relying on any model-dependent assumptions. Moreover, our results are in excellent agreement with numerical predictions~\cite{Ec_Delande2017}, highlighting the potential of this new scheme and opening promising avenues for quantitative studies of quantum criticality in disordered systems.

\section*{Experimental scheme}

\begin{figure} 
	\centering
	\includegraphics[width=0.45\textwidth]{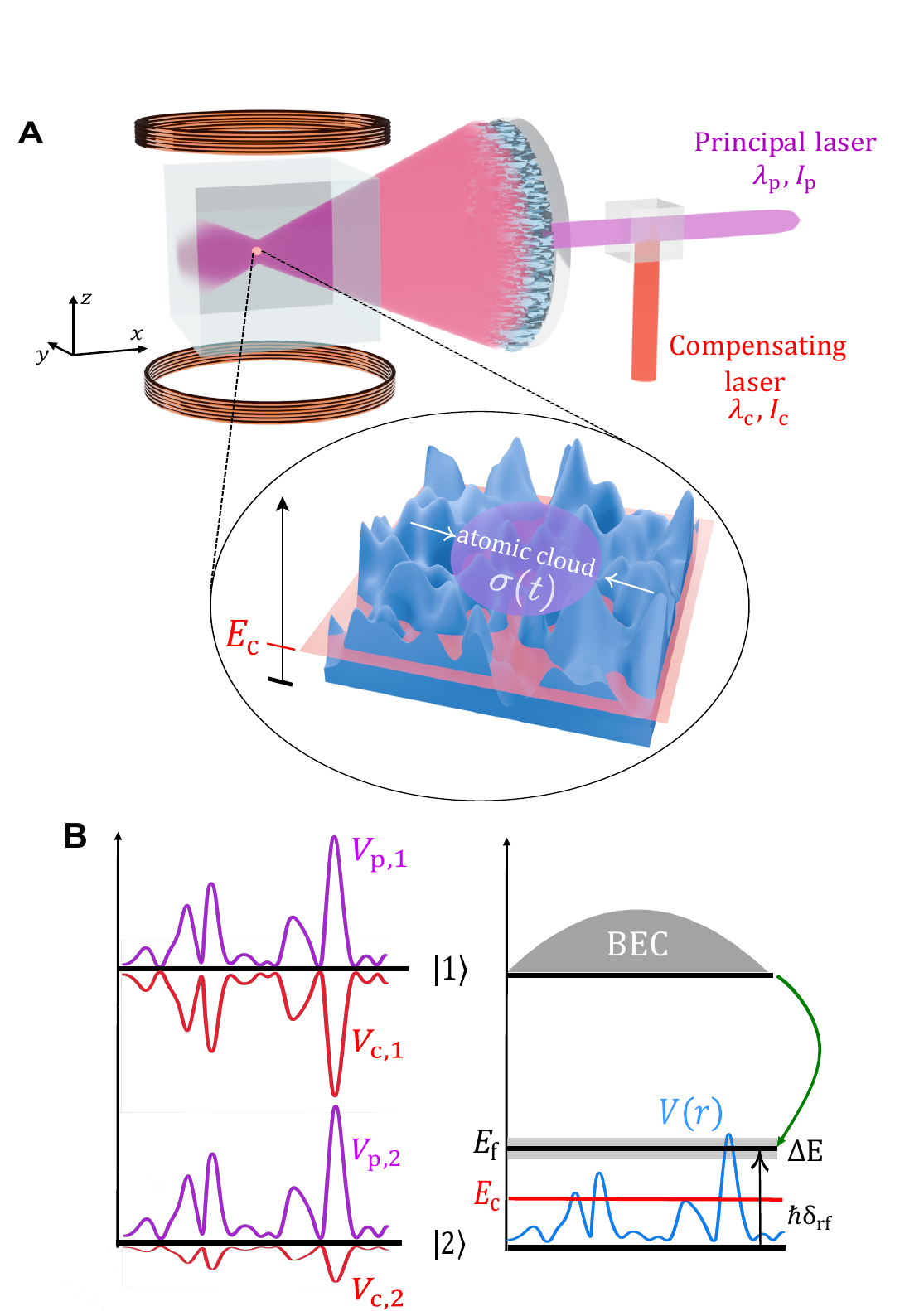} 
		\caption{\textbf{Energy-resolved characterization of 3D Anderson transition.} (\textbf{A-B}) Experimental scheme: a $^{87}$Rb Bose-Einstein condensate (BEC), initially prepared in the disorder-free state $|1\rangle$, is partially transferred via two-photon rf transition into the disorder-sensitive state $|2\rangle$ with a well-defined energy $\Ef$ and a narrow width $\Delta E=h/t_\text{rf}$, with $t_\text{rf}$ the rf pulse duration. Both states are suspended against gravity using a magnetic levitation (yellow coils). The state-dependent disorder is achieved by using two laser speckle fields, a principal (purple) and a compensating (red) one. Their wavelengths and amplitudes are finely tuned to cancel the disorder in state $|1\rangle$, and not in the state $|2\rangle$ (see Methods). Inset of (\textbf{A}): Schematic energy spectrum in the disordered speckle potential $V$, illustrating the position of the mobility edge $E_c$. With our setup, we probe the dynamics of the atomic sample through its cloud size $\sigma(t)$. The expansion is diffusive for $E_\text{f}>E_c$, $\sigma(t)\propto \sqrt{t}$, whereas its saturates for $E_\text{f}<E_c$, $\sigma(t)=\text{const}$. At the mobility edge $E=E_c$, finally, the dynamics is subdiffusive, with the critical behaviour $\sigma(t)\propto t^{1/3}$. }
	\label{fig:scheme}
\end{figure}

Our experimental characterization of the Anderson transition is based on an experimental protocol allowing us to load the atoms in the disorder at a well-defined energy: A fraction of a $^{87}$Rb Bose-Einstein Condensate (BEC) is transferred from a disorder-free state~$|1\rangle$ into a disorder-sensitive state~$|2\rangle$ at a chosen energy $\Ef$, using a radio-frequency (rf) pulse (Fig.~\ref{fig:scheme}). In the weak-coupling regime, this process is well described as the coupling of a discrete state to a quasi-continuum, where the transfer rate $\Gamma$ is estimated using Fermi's golden rule~\cite{Volchkov2018,lecoutre2022bichromatic}. For short interaction times
$\Gamma t_\mathrm{rf} \ll 1$, with $t_\mathrm{rf} $ the rf-pulse duration, the width $\Delta E$ of the energy distribution of the transferred atoms is Fourier-limited, with $\Delta E\sim h/  t_\mathrm{rf}$.
This method provides a unique control of the atom's energy in the disordered potential, where the target energy $\Ef$ is adjusted through the rf frequency to probe the Anderson transition with high resolution. The technique, originally demonstrated in~\cite{Volchkov2018,lecoutre2022bichromatic} for spectral-function measurements, has so far been limited by atom losses and heating arising from two-body spin collisions in the disordered state~\cite{XudongPHDThesis}. In the present work, we implement an improved scheme to generate the state-dependent disorder with an increased lifetime up to 5~seconds, using the states $|1\rangle = |F=2,\,m_F=1\rangle$ and $|2\rangle = |F=1,\,m_F=-1\rangle$ (see Methods). These ``clock states" exhibit equal magnetic susceptibility at the ``magic" magnetic field of $B^\star$~=~\SI{3.23}{G}, a property that we exploit to suspend both states against gravity using magnetic levitation and to perform rf transfers immune to magnetic noise. These advances allow us to monitor atomic propagation in the disordered potential for several seconds, a crucial capability for discriminating between localized and diffusive dynamics.
\begin{figure*} 
		\centering
		\includegraphics[scale=1.35]{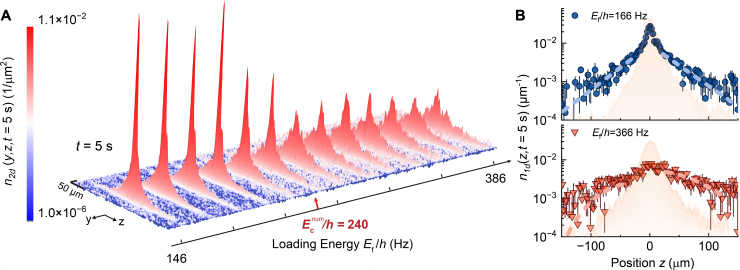} 
		\caption{\textbf{Direct observation of the Anderson transition.} (\textbf{A}) Two-dimensional column density in the $(y,z)$ plane, recorded after $\SI{5}{\second}$ of expansion in a speckle potential of amplitude $\VR/h=$~\SI{416}{Hz}. It shows a striking change in the behavior of the central density depending on the loading energy $\Ef$. For proper comparison, all density profiles are normalized to a fixed number of atoms. The numerically predicted position of the mobility edge from Ref.~\cite{Ec_Delande2017}, $\Ec^\text{num}/h=240$~Hz, is indicated on the energy axis. (\textbf{B}) Integrated one-dimensional atomic densities along the $z$ axis, $n_\mathrm{1d}(z,t=\SI{5}{\second})$, displayed on a semi-log scale and corresponding to the column densities shown in (\textbf{A}) for the low energy $\Ef/h=\SI{166}{Hz}$ (blue plot) and the high energy $\Ef/h=\SI{366}{Hz}$ (red plot). The light orange shading indicates the initial density profile. The wings of the profiles are independently fitted with an exponential (blue) or a Gaussian (red) function.}
		\label{fig:Transition}
	\end{figure*}

We first prepare a very dilute BEC of about $2\times10^5$ atoms in state $|1\rangle$, suspended against gravity and confined in a shallow and nearly isotropic optical trap. A low temperature around $T\sim~$\SI{7}{\nano K} is reached, yielding an almost pure BEC,  with 75\% condensate fraction, a chemical potential $\mu_\text{in}/h\sim$~\SI{350}{Hz} and a geometric mean Thomas-Fermi radius~$\overline{R} \sim~$\SI{14}{\micro\meter} (see Supplementary Information). 
The 3D state-dependent disordered potential is generated by superimposing two speckle beams near 780~nm, with a slight wavelength difference ($\delta \lambda/\lambda \sim 2.5 \times 10^{-4}$) -- see Methods. 
The speckle patterns, referred as the principal and the compensating one, are  almost identical for both lasers, as illustrated in Fig.~\ref{fig:scheme}. They are created by shining two beams combined in a single mode fiber through the same diffuser and by focusing the light on the atoms. The relative intensities and detunings of the two lasers are adjusted such that the state $|1\rangle$ is essentially disorder-free, while $|2\rangle$ experiences a finite disorder potential of rms amplitude $\VR$. The speckle disorder is ``repulsive", meaning that intensity maxima correspond to potential maxima for atoms in state $|2\rangle$. Noticeably, the classical percolation threshold is extremely low for such laser speckle disorder (it almost corresponds to the bottom energy $E=0$) so that the following observations cannot be interpreted by simple classical trapping~\cite{Fred2012}. Spatially, the 3D speckle pattern is elongated along the $x$ axis, with an averaged correlation length $\overline{\sigma}=0.5\,\mu$m, defined as the geometric mean along each direction (see Methods). This length sets the correlation energy $E_{\sigma}=\hbar^2/m \overline{\sigma}^2\approx h \times \SI{441}{Hz}$, an important energy scale that plays a central role in determining the transport properties of atoms in disorder~\cite{Kuhn_2007}. Most importantly, the normalized disorder amplitude $\eta= \VR/E_{\sigma}$ enables mobility-edge predictions obtained from different speckle geometry to collapse onto a single ``universal" curve~\cite{Ec_Delande2017}.

We transfer a small fraction -- around 5$\%$ -- of the atoms using a two-photon rf pulse of typical duration $\trf=\SI{40}{ms}$. Its temporal envelope, a Kaiser function, is chosen to minimize side lobes in the energy distribution, leading to an energy rms width estimated to $\Delta E/h=\SI{14}{Hz}$ (see Methods). This narrow energy width, a key asset of the present work, represents a reduction of one~\cite{Fred2012,G.Semeghini2015} or even two~\cite{kondov2011three} orders of magnitude compared with previous experiments. The total rf frequency (accounting for the two rf photons) sets the targeted energy to $\Ef=\hbar \delta_\mathrm{rf}$, where $\delta_\mathrm{rf}$ corresponds to the detuning from the bottom energy level $E=0$ of the disorder, as illustrated in Fig.~\ref{fig:scheme}. Immediately after the rf transfer, we switch off the optical trap at $t=0$ and remove the atoms from state $|1\rangle$. We then monitor the subsequent evolution of the atoms in state $|2\rangle$ in presence of the disordered potential. After a given expansion time $t$, the \emph{in-situ} column density $n_{\rm col}(y,z,t)=\int n(x,y,z,t){\rm d}x$, where $n(x,y,z,t)$ denotes the full 3D atomic density, is recorded by fluorescence imaging along the $x$ axis. To allow for a proper comparison of the density profiles and to eliminate the effect of spatially homogeneous atom loss, the column density is further normalized to unit integral. We then systematically investigate the temporal dynamics for various disorder amplitudes $\VR$ and energies $\Ef$.

\begin{figure*}
		\centering
		\includegraphics[scale=0.85]{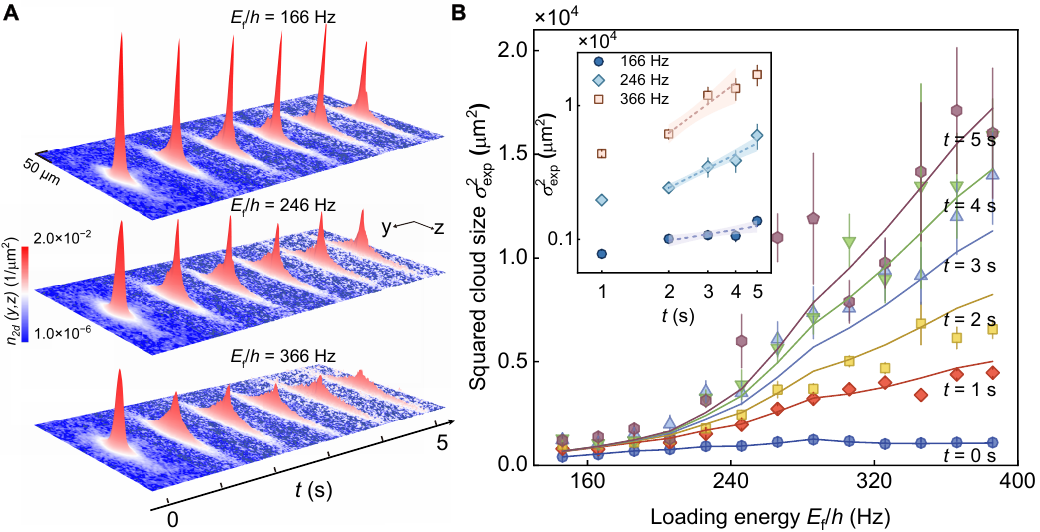}
		\caption{\textbf{Dynamical behaviour as a function of the energy $\Ef$.} (\textbf{A}) Experimental column-density profiles as a function of time for three fixed energies $\Ef$ (166, 246 and 366 Hz) for the disorder amplitude $\VR/h=$~\SI{416}{Hz}. As in Fig.~\ref{fig:Transition}, the profiles are normalized to a fixed number of atoms. The initial cloud, which corresponds to atoms transferred from the same BEC in state $|1\rangle$ but with a different rf frequency, may slightly vary with the energy $\Ef$ (see Supplementary Information), see also panel B. (\textbf{B}) Main panel: squared size $\sigma^2_\text{exp}(t)$ of the cloud (symbols) as a function of loading energy $\Ef$ for different expansion times (same disorder amplitude $\VR/h=$~\SI{416}{Hz}). Solid lines show theoretical predictions based on the 3D self-consistent theory of Anderson localization adapted to our configuration (see Methods). Inset: Time evolution, in a log-log plot, of the squared cloud size $\sigma^2_\text{exp}(t)$ at the three energies $\Ef$ corresponding to the profiles in (\textbf{A}). The dashed lines show fits of the form $\sigma^2_\text{fit}(t)=A t^\kappa$, performed on data with $t>$ \SI{1}{s}. The colored bands denote one-standard-deviation confidence intervals. }
		\label{fig:Dynamics}
	\end{figure*}

\section*{Direct observation of the 3D Anderson transition}

Fig.~\ref{fig:Transition}A shows the measured column-density profiles for increasing loading energies $\Ef$ after $t=\SI{5}{s}$ of evolution in a speckle potential of amplitude $\VR/h=416$ Hz, corresponding to the normalized disorder strength $\eta=0.94$. A striking change in behaviour is observed as the energy is increased: while the profiles keep the same peaked structure below $\SI{206}{Hz}$, the central atomic density decreases significantly as the cloud becomes markedly broader once the energy exceeds $\SI{266}{Hz}$. In between, we observe a smooth crossover occurring precisely around the mobility edge numerically predicted in Ref.~\cite{Ec_Delande2017}, $E_\mathrm{c}^\mathrm{num} /h=\SI{240\pm 6}{Hz}$. As detailed below, this behaviour constitutes nothing else than the direct observation of the 3D Anderson transition.

Before going further, several important remarks must be emphasized. First, the Anderson transition appears in Fig.~\ref{fig:Transition}A as a crossover rather than a sharp transition. As for any second-order phase transition, this stems from the fact that the transition becomes strictly sharp only in the ``thermodynamic'' limit of infinitely long evolution times. At finite time, the transition is therefore inevitably broadened -- even in the ideal case of a perfect energy selection with $\Delta E=0$. 
Second, the density profiles result from averaging images over 8 to 9 experimental runs under the same conditions. Such averaging is used to increase the signal-to-noise ratio due to the low atom numbers transferred in state $|2\rangle$. 
Additional self-averaging arising from the finite imaging resolution and energy width further suppresses residual disorder-induced fluctuations~\cite{Volchkov2018}. Last, in our experimental configuration, a weak residual harmonic confinement of frequency $\omega_\perp/2\pi=\SI{7}{\hertz}$ persists in the horizontal $x-y$ plane due to magnetic levitation (see Methods). Given the very low energy of the atoms, this confinement is sufficient to suppress their transverse expansion, resulting in a typical rms transverse size of about $\SI{10}{\micro \meter}$. 
This size remains much larger than the scattering mean free path $\ell\simeq \SI{1.5}{\micro\meter}$, which we estimate using our previous direct measurement of the scattering time $\tau\sim \SI{1}{\milli \second}$~\cite{richard2019TauS} (see Supplementary Information). This ensures that the system retains a genuinely three-dimensional geometry. This conclusion is further supported by the excellent agreement between our measurements of the mobility edge and the numerical predictions~\cite{Ec_Delande2017}, as presented below.

To proceed, we analyze the decay of the atomic density profiles. Since the expansion occurs predominantly along the $z$ axis, we henceforth consider the integrated 1D density profiles $n_{1\rm d}(z,t)=\int n_{\rm col}(y,z,t) {\rm d}y $, which are shown on a semi-logarithmic scale in Fig.~\ref{fig:Transition}B for two different energies. At low energy, $\Ef/h=\SI{166}{\hertz}$ (blue plot), the density tails exhibit a clear exponential decay -- a hallmark of Anderson localization. In contrast, at higher energy $\Ef/h=\SI{366}{\hertz}$ (red plot), the profile is significantly broader and flatter, and is compatible with a Gaussian shape, as expected in the diffusive regime.
While the observed profiles are fully consistent with the 3D Anderson transition picture, the present analysis alone does not provide a definitive proof. This is partly due to the low signal-to-noise ratio in the density tails, especially in the diffusive regime, where the matter wave has significantly spread, and to the residual energy broadening $\Delta E$, which may alter the tail behavior compared to the ideal ``monochromatic" case~\cite{Palencia2007,Bourdel2010,muller2014comment}.

A more robust signature of the Anderson transition is obtained by examining the temporal evolution of the cloud size $\sigma(t)$. Above the mobility edge $\Ef>\Ec$, the dynamics is diffusive such that the squared size grows linearly in time ($ \sigma^2 \propto t$). Below the mobility edge, $\Ef<\Ec$, propagation is halted by Anderson localization and $\sigma^2(t)$ saturates at long time. At the mobility edge, $\Ef=\Ec$, the scaling theory of the Anderson transition predicts a critical anomalous diffusion, with $\sigma^2 (t) \propto t^{2/3}$~\cite{Ohtsuki1997, abrahams1979}. As shown in Fig.~\ref{fig:Dynamics}A, qualitatively different behaviours are indeed visible for the cloud dynamics depending on the energy.

To characterize  the cloud size $\sigma(t)$, one could in principle estimate the rms width of the density profiles. However, this quantity is highly sensitive to measurement noise in the low-density tails. Instead, we  use the inverse of the one-dimensional integrated central density as a more robust estimator of the cloud size, following the approach of Ref.~\cite{chabe2008}. We thus define $\sigma_\mathrm{exp}(t)=1/n_{1\rm d}(z=0,t)$, and plot its squared value in Fig.~\ref{fig:Dynamics}B for the dataset corresponding to $\VR/h=\SI{416}{Hz}$. The overall behaviour as a function of energy $\Ef$ and for increasing evolution times confirms the conclusions drawn above from the profile analysis.
First, the nearly frozen dynamics persisting over several seconds provides additional evidence of localization at low energies, for $\Ef /h\leq \SI{206}{Hz}$. Second, a crossover is again observed, with an increasingly rapid expansion for energies $\Ef /h\geq \SI{226}{Hz}$. In particular, a marked expansion occurs around the theoretically predicted mobility edge $\Ec /h = 240$~Hz, consistent with the expected critical anomalous diffusion.

To further support our observations, we also computed $1/n_\text{1d}^2(z=0,t)$ using the 3D time-dependent self-consistent theory of Anderson localization~\cite{Vollhardt1980, Cherroret2025}. The calculation takes into account the initial shape of the atomic cloud, the finite energy width $\Delta E$, and the experimentally estimated value of the mobility edge (see Methods and Supplemental Information). The theoretical results, shown as solid curves in Fig.~\ref{fig:Dynamics}B, are in good agreement with the experimental data.

\section*{Determination of mobility edge}
	
	\begin{figure}
			\centering			
            \includegraphics[scale=0.55]{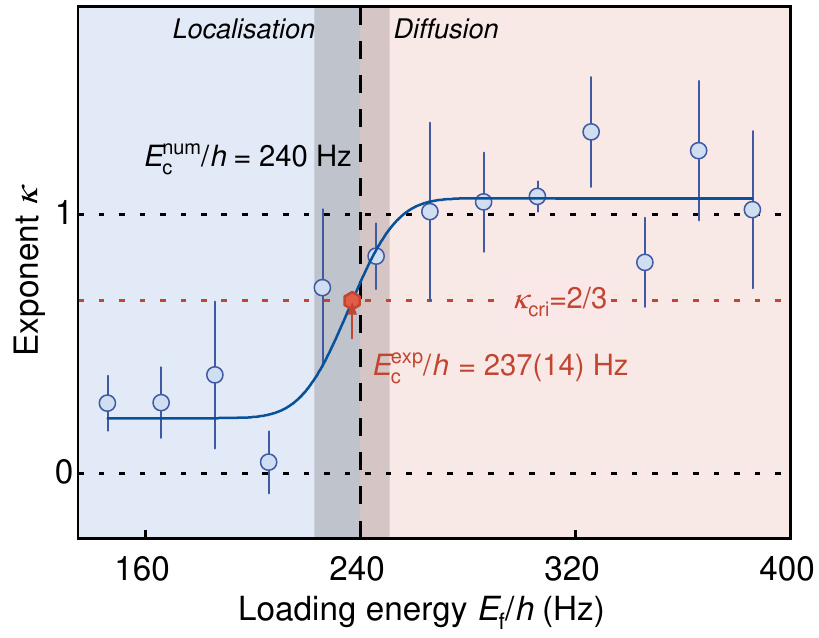} 
			\caption{\textbf{Direct estimation of the mobility edge for $\VR/h=$~\SI{416}{Hz} ($\eta=0.94$).}
            Exponent $\kappa$, extracted from fits of the squared cloud width to $\sigma^2_\text{exp}(t)=A t^\kappa$ (see inset of Fig.~\ref{fig:Dynamics}B), shown  as a function of loading energy $\Ef$, for the disorder amplitude $\VR/h=$~\SI{416}{Hz}.  The red dotted horizontal line indicates the critical value $\kappa_\text{cri}=2/3$, corresponding to the subdiffusive dynamics expected at the mobility edge, while the two grey dotted horizontal lines mark the localized ($\kappa=0$) and diffusive ($\kappa=1$) limits. 
            The blue solid curve represents the \emph{ad-hoc} fitted function (an error function, see Methods). The shaded area represents the experimentally inferred critical region, whose half-width is defined by the rms width $W_{\rm exp}$ of the fitting function. The vertical black dashed line indicates the numerically predicted mobility edge from Ref.~\cite{Ec_Delande2017}, while the red point marks the experimentally estimated mobility edge $\Ec^{\rm exp}$.}
			\label{fig:ME_measure}
		\end{figure}

To accurately determine the mobility edge, we analyze in detail the temporal dynamics of the cloud size and examine the associated scaling behaviour with energy. As an example, the inset of Fig.~\ref{fig:Dynamics}B shows representative time evolutions of $\sigma^2_\text{exp}(t)$ for three different energies that span across the theoretically predicted mobility edge, and corresponding to the profiles in Fig.~\ref{fig:Dynamics}A. By fitting the data to the power-law $ \sigma_\mathrm{exp}^2(t)= At^{\kappa}$, using only times  $t > \SI{1}{s}$ to minimize finite-time effects, we extract the wave-packet growth exponent $\kappa$ as a function of the loading energy $\Ef$. The results are plotted in Fig.~\ref{fig:ME_measure}, where the three ideal behaviours (localized, critical, and diffusive) are indicated as horizontal dashed lines. We observe that $\kappa$ evolves continuously from a value near zero at low energy, characteristic of the localized regime, to a value approaching unity at large energy, consistent with diffusive spreading. We also note that $\kappa$ does not fully vanish at low $\Ef$, as would be expected for perfectly energy-resolved states deep in the localized regime. Although the precise origin of this residual value is unclear, we attribute it to a combination of the finite energy width of the energy selection -- allowing a fraction of the energy distribution to exceed the mobility edge -- and residual experimental excitations (see Discussion).

To obtain a quantitative estimate of the mobility edge, we fit the data in Fig.~\ref{fig:ME_measure} with an \emph{ad-hoc} fitting ``S-curve" function (an error function -- see Methods) shown as a solid line. We then define the experimental mobility edge as the energy at which the fitting function intersects the horizontal line $\kappa_{\rm cri} = 2/3$, corresponding to the critical scaling. This procedure yields $E_\mathrm{c}^\mathrm{exp}/h = \SI{237(12)}{Hz}$ in excellent agreement with the numerically predicted value $E_\mathrm{c}^\mathrm{num}/h=\SI{240(6)}{Hz}$ mentioned above. The fit also provides $W_\mathrm{exp}  = \SI{14}{Hz}$ for the rms width of the critical region, illustrated as the gray shaded area in Fig.~\ref{fig:ME_measure}. This width is compatible with the finite energy spread $\Delta E$ of the atomic cloud, but it also accounts for the fundamental width of the critical region associated to our finite-time analysis. Consistently, the estimated width $W_\mathrm{exp}$ is not too far from the theoretical prediction $W_\mathrm{c} \simeq \eta^4 E_\sigma(\tau/t)^{0.21}/(\sqrt{2\ln 2}h) \simeq 49$ Hz reported in Ref.~\cite{muller2016critical}, using the scattering time $\tau\simeq\SI{1}{\milli \second}$~\cite{richard2019TauS}, even though this expression strictly applies in the limit $\eta\ll1$.

        	\begin{figure}
			\centering			
			\includegraphics[scale=0.55]{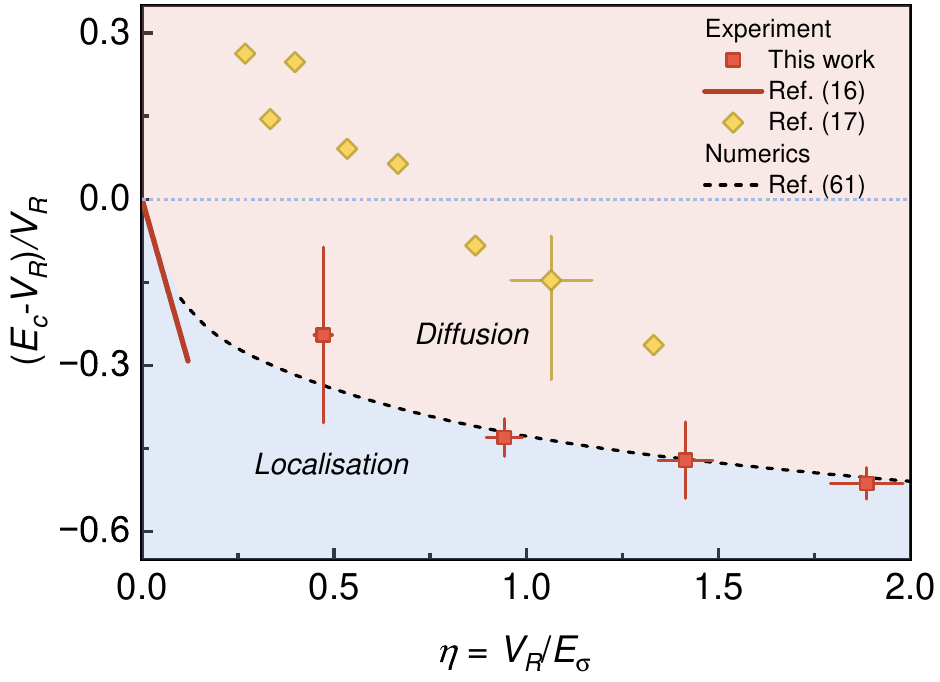} 
			\caption{\textbf{State of the art: measurements of the mobility edge and comparison with numerics.}
            Mobility edge $\Ec$, relative to the disorder mean value $\VR$ (horizontal dotted line), shown as a function of the normalized disorder strength $\eta = \VR/E_\sigma$, where $E_{\sigma}$ is the correlation energy. The red squares show the direct measurements of $\Ec^\mathrm{exp}$ obtained in the present work at four disorder amplitudes, using the methodology presented in Fig.~\ref{fig:Dynamics}. The vertical error bars indicate the half-widths $W_{\rm exp}$ of the critical regions, while the disorder amplitudes are calibrated with a 5$\%$ uncertainty (horizontal error bars) -- see supplementary Information. The thick red line correspond to the estimated scaling of the mobility edge estimated in our previous work~\cite{Fred2012}, resulting from a comparison between experiments and theoretical modeling. The yellow diamonds indicate the experimental estimation of the mobility edge from Ref.~\cite{G.Semeghini2015}. Last, the black dashed curve corresponds to the numerical prediction of Ref.~\cite{Ec_Delande2017} for repulsive laser speckle disorder. 
			}
			\label{fig:MEcomparison}
		\end{figure}

By repeating the same analysis for different disorder strengths $\VR$, we experimentally determine the evolution of the mobility edge. The results are summarized in Fig.~\ref{fig:MEcomparison}, which shows the normalized mobility edge $(\Ec^\mathrm{exp} - \VR)/\VR$, relative to the mean potential, as a function of the normalized disorder strength $\eta = \VR/E_\sigma$. For each value of $\eta$, the vertical error bars indicate the rms width $W_\mathrm{exp} $ of the critical region. Our measurements are in excellent agreement with the numerical prediction of Ref.~\cite{Ec_Delande2017}, except at the lowest disorder amplitude ($\eta \sim 0.5$). This regime indeed corresponds to very low energy scales given our correlation energy $E_\sigma$. The experiment then becomes extremely sensitive to residual perturbations, leading to larger uncertainties.
For comparison, the yellow diamonds and green solid line indicate, respectively, the experimental estimates reported in Ref.~\cite{G.Semeghini2015} and those from our previous work~\cite{Fred2012}. 
We note that the latter relied on an assumed quadratic scaling of the mobility edge at weak disorder,
$(\Ec^\mathrm{exp} - \VR) \propto  \beta \VR^2/E_\sigma$, with $\beta\sim -2.44 $ determined from a combination of experimental observations and theoretical modeling 
(in the representation of Fig.~\ref{fig:MEcomparison}, this quadratic dependence appears as a linear behavior). The result of Ref.~\cite{kondov2011three} is not shown, as it lies far above the vertical range of the figure (see Ref.~\cite{Ec_Delande2017}). Note that the classical percolation threshold $E_{\rm cl}$ almost coincides with the bottom of the energy spectrum, $(E_{\rm cl}-\VR)/\VR\simeq-1$, and thus lies far below the observed mobility edge.

\section*{Discussion and outlook}

The measurements presented in Fig.~\ref{fig:MEcomparison} provide a direct determination of the mobility edge for the 3D Anderson transition, based only on experimental observation. 
This achievement relies on our ability, thanks to an original rf transfer method, to resolve the matter-wave dynamics on both sides of the transition. This is in sharp contrast to prior experiments that used strongly energy-broadened atomic clouds, where the mobility edge could only be inferred indirectly. 
As expressed as a function of the dimensionless disorder amplitude $\eta= \VR/E_{\sigma}$, the mobility-edge measurements reported in Fig.~\ref{fig:MEcomparison} apply to any repulsive speckle disorder, independently of the specific value of the correlation length or of the anisotropy. Their excellent agreement with numerical predictions further demonstrates that our approach constitutes a significant advance for the study of Anderson transitions and mobility edges with ultracold atoms in real space disorder.

Several aspects of the present work, however, call for further clarification or improvement in future experiments.
First, the precise impact of the shallow confinement in the transverse $x-y$ plane -- induced by the magnetic levitation -- on the critical regime remains to be fully understood. Although no significant effect has been observed so far, a dedicated theoretical analysis would be desirable. 
Second, the role of the small, though not entirely negligible, thermal fraction in the initial atomic cloud warrants clarification. This residual thermal component could indeed give rise to a weak high-energy tail in the atomic energy distribution, thereby affecting the density profiles in the outer wings. 
Last, residual atom-atom interactions, which cannot be completely eliminated, may also play a role. These interactions are predicted to partially destroy localization over long times~\cite{Kopidakis2008,Pikovsky2008,Cherroret2014, Scoquart2020}. Experimentally, we mitigate this effect by using very low atom numbers. 
Based on the initial density profile, we estimate a very weak interaction energy at $t=0$,  $E_{\mathrm{int}}/h \approx \SI{5}{Hz}$, which decreases even further during the subsequent expansion. Nevertheless, higher local densities -- unresolved by our imaging setup -- may be reached when loading the atoms into the localized regime, potentially leading to a local increase of the interaction energy and a slight broadening of the energy distribution.
Altogether, we tentatively attribute the observed residual $\kappa$ exponent in the localized regime (Fig.~\ref{fig:ME_measure}) to a combination of residual temperature and interactions effects. These questions are left open for future investigations.

The above-mentioned limitations are essentially technical and will likely be overcome in future work, paving the way for a quantitative exploration of universal critical phenomena at the transition. This includes the determination of the critical exponent~\cite{abrahams1979} and the study of wave-function multifractality at the critical point~\cite{mirlin2000,mirlin2000statistics,Werner2018, akridas2019multifractality}. Combined with our precise control over disorder, the technique also provides a powerful platform to investigate how interactions modify the mobility edge in dilute gases~\cite{Cherroret2014,Cherroret2021}, and to assess whether many-body mobility edges emerge -- or fail to emerge -- in the many-body localization scenario \cite{Luitz2015,Roeck2016}.


\vspace{0.5cm}

\section*{Methods}

{\it State-dependent disorder ---}
The principal and the compensating lasers are respectively detuned by $\Delta _p/2\pi\approx$\SI{100}{GHz} and $\Delta _c/2\pi\approx$\SI{-9,1}{GHz} from the $D_2$ line of $^{87}\mathrm{Rb}$ ($\lambda_L$=\SI{780.23}{nm}). The lasers intensities are adjusted to achieve an almost perfect cancellation in the  state $|1\rangle $ and a finite disordered potential of rms amplitude $\VR$ in state $|2\rangle $:   $V_{\mathrm{p},1} (\mathbf{r}) + V_{\mathrm{c},1} \approx 0$ and  $V_{\mathrm{p},2} (\mathbf{r}) + V_{\mathrm{c},2} (\mathbf{r})=V(\mathbf{r})$ (see Fig.~\ref{fig:scheme}). For a speckle disorder, the amplitude distribution follows an exponential law, such that the rms amplitude corresponds to the mean value, that is $\VR=\langle V \rangle$. The photon scattering induced lifetime is increased by ramping down to zero the compensating laser in \SI{10}{\milli \second} at $t=0$. The principal laser intensity is adjusted in parallel to maintain the disorder amplitude $\VR$ constant. For $\VR/h=\SI{416}{\hertz}$, we measure a lifetime around $\SI{5.0(5)}{\second}$. More details are given in the Supplementary Information. 
\\

{\it Laser speckle geometry ---}
The laser speckle fields are created by illuminating a diffuser of diameter \SI{20.3(1)}{mm} with identical Gaussian beams of waist  $w=\SI{9(1)}{mm}$ propagating along the $x$ axis, and by focusing the transmitted light onto the atoms at a distance of \SI{15.2(5)}{mm}. This geometry corresponds to the numerical-aperture $\mathrm{NA}=0.55(2)$. The correlation lengths are extracted from the measurement of the  two-point intensity correlation function~\cite{Volchkov2018}, following the convention  of Ref.~\cite{Ec_Delande2017}. It yields $\sigma_{\perp} = \SI{0.30}{\micro \meter}$  in the $y-z$ plane and $\sigma_{\parallel}= \SI{1.45}{\micro \meter}$ along the $x$ axis.
\\

{\it Two photon rf transfer ---}
 The transition $|1\rangle\rightarrow|2\rangle$ is driven by a two-photon rf-process involving the intermediate state $|0\rangle=|F=2,m_F=0\rangle$. A first microwave field of frequency $\omega_1/2\pi \sim \SI{6.83}{\giga \hertz}$ couples the $|1\rangle\rightarrow|0\rangle$ with a detuning  $\delta_0/ 2\pi\sim\SI{0.5}{\mega \hertz}$.  A second rf field of frequency $\omega_2 /2\pi  \sim \SI{2.8}{\mega \hertz} $ then drives the transition $|0\rangle\rightarrow|2\rangle$. The couplings are weak enough such that the system reduces to an effective single rf transition from  $|1\rangle\rightarrow|2\rangle$, driven at the total frequency $\omega_{\rm rf}=\omega_1+\omega_2$, with the effective Rabi frequency $\Omega_{\rm rf}=\Omega_1\Omega_2/2\delta_{0}$ ($\Omega_{1,2}$ refer to the Rabi frequencies of each fields). The loading energy is set by $E_{\rm f}=\hbar \delta_{\rm rf}$, where $\delta_{\rm rf}= \Delta_{\rm HFS} - \omega_{\rm rf}$ is the detuning from the bare resonant frequency (that is, in the absence of disorder) corresponding to the hyperfine splitting between the respective internal energies $\Delta_{\rm HFS}/2\pi \sim \SI{6.8}{GHz}$. The rf pulse shape is a Kaiser window function given by:
\begin{equation}
	f(t)=\frac{I_0\!\left[2 \pi \sqrt{1-\left({2t}/{t_{\mathrm{rf}}}-1\right)^2}\right]}{I_0(2\pi)}, 
	\qquad 0<t< t_{\mathrm{rf}} \; \nonumber,
\end{equation}
For $t_\mathrm{rf} = 40$~ms, the rms width of the Fourier-limited energy width is $\Delta E /h \simeq \SI{13}{\hertz}$ (see Supplementary information for details). 
 \\

{\it Magnetic levitation ---}
The magnetic levitation is created by combining a vertical magnetic field gradient $b^\prime \sim m g/ m_F g_F \mu_{\rm B} \sim  \SI{30.5} \, \mathrm{G/cm}$ ($g_F$ is the Land\'e factor and $ \mu_{\rm B}$ the Bohr's magneton), together with an homogeneous vertical bias field $B^\star=\SI{3.23}{G}$. The magnetic fields are created with coils in the (anti-) Helmholtz configuration, yielding no residual confinement along the vertical axis ($\omega_z\approx 0$). This configuration yields a residual harmonic confinement $\omega_\perp =  \sqrt{m g^2 / 4 m_F g_F \mu_{\rm B} B^\star} \approx 2\pi\times 7~\mathrm{Hz}$~\cite{sackett2006limits}, a value confirmed experimentally.
\\

{\it Fluorescence imaging ---}
The atomic density distribution is recorded using a high-sensitivity electron-multiplying CCD (EMCCD) camera. A saturating resonant probe beam is applied along the $y$ axis for a duration of \SI{20}{\micro\second}. The fluorescence light is collected around the $x$ axis with a numerical aperture ${\rm NA}\sim 0.5$. The imaging resolution is set by the effective pixel size of \SI{2.4}{\micro \meter} along the $z$ vertical axis. 
\\

{\it 1D central density estimation ---}
The 1D central density $n_{1\rm d}(z=0,t)$ is determined by (i) normalizing the 1D integrated profile  $n_{1\rm d}(z,t)$ to unity and (ii) fitting the vicinity of the cloud center $z\sim0$ with an inverted parabola. This procedure yields a more reliable estimate of the atomic density at the cloud center. The error bars for $\sigma_{\rm exp}$ account for both statistical uncertainties and uncertainties arising from the fit. Note that the determination of the central density is subject to a high degree of uncertainty at $t=\SI{5}{\second}$, due to the very low atomic density, especially in the diffuse regime where it has even further decrease due to the expansion. Thus, we discard the $t=\SI{5}{\second}$ data for $E_{\rm f}/h \geq \SI{286}{\hertz}$ -- that is when the diffusive behaviour is well established --  for the extraction of the exponent $\kappa$.
\\

{\it Estimation of the mobility edge ---}
We fit the dynamical exponents $\kappa$ with the empirical error function  $ a \left[ \mathrm{erf}\left( (E_{\text{f}} - \bar E)/(\sqrt{2}W_{\rm exp} \right) + 1 \right] / 2 + b $, $a$, $b$, $\bar E$ and $W_{\rm exp}$ being free parameters. The mobility edge  $E_c$ is then obtained by identifying the energy at which the fitted curve reaches the critical value $\kappa_{\rm cri}=2/3$.
\\

{\it Self-consistent theory ---}
We use the self-consistent theory of localization to compute the 1D density  
$n_{1d}(z,t)=\int dz' dE' \mathcal{D}(E';E_f)P_{E'}(z,z',t)n_\text{1d}(z',0)$, where  $\mathcal{D}(E';E_f)$ is the estimated energy distribution centered around the loading energy $\Ef$, and $n_{\rm 1D}(z',0)$ is the  experimentally measured 1D density profile. The propagator 
\begin{equation}
 P_{E}(z,z',t) = \int_{-\infty}^\infty \frac{d\omega}{2\pi}\int_{-\infty}^\infty \frac{dq}{2\pi}\frac{e^{iq(z-z')-i\omega t}}{-i\omega+D(\omega)q^2} \; .
\end{equation}
is evaluated numerically. The generalized, frequency-dependent diffusion constant $D(\omega)$  is obtained from the self-consistent theory through
\begin{align}
\label{eq:SCTL_D_Method}
    \frac{1}{D(\omega)}\!=\!\frac{1}{D_B} +\frac{1}{\pi\rho\hbar D_B}
    \int^{Q_\text{UV}}\!\! \frac{d^3\boldsymbol{Q}}{(2\pi)^3}
    \frac{1}{-i\omega+D(\omega)\boldsymbol{Q}^2}.
\end{align}
$D(\omega)$ depends on three independent parameters, the classical diffusion constant $D_B$, the 3D density of state per unit volume $\rho$, and a momentum cutoff $Q_\text{UV}$ regularizing the theory at short length scales. These parameters are recast into three composite parameters: $\alpha_0$, $\beta$, and the mobility edge $E_c$. The latter is fixed to its experimental estimation $\Ec^{\rm exp}/h$, while $\alpha_0, \beta$ are treated as free parameters and optimized to achieve the best overall agreement between theory and experiment across the full data set (see Supplemental Information).
\\



\textbf{Acknowledgments ---}
The authors thank V. Denechaud, B. Lecoutre, A. Signoles for early work and D. Clément, D. Delande and M. Filoche for fruitful discussions. This work benefited financial support from the project Localization of Waves of the Simons Foundation (Grant No. 601939), the French ANR under Grant No. ANR-24-CE30-6695 (FUSIoN) and Region Île-de-France in the framework of DIM QuanTiP.
~\\


\textbf{Data availability ---}
The datasets generated and analysed in the present study are available from the corresponding author upon reasonable request.

\bibliographystyle{naturemag}
\bibliography{3DAL_bib_arXiv}


\clearpage
\newpage
\makeatletter
\renewcommand{\thefigure}{S\arabic{figure}}
\renewcommand{\theequation}{S\arabic{equation}}

\renewcommand{\theHfigure}{S\arabic{figure}}
\renewcommand{\theHequation}{S\arabic{equation}}
\makeatother
\setcounter{figure}{0}
\setcounter{table}{0}
\setcounter{section}{0}
\setcounter{equation}{0}
\appendix

\makeatletter
\let\oldtitle\@title
\title{Supplementary information for: \\ \oldtitle}
\makeatother

\maketitle

\subsection{State-dependent disorder  \label{Sec:StateDependent}}

A key feature of the experimental scheme is the use of a state\textcolor{blue}{-}dependent disorder. The creation of state dependent potentials for alkali atoms has been widely investigated in the context of optical lattices using circularly polarized light tuned between the $D_1$ and $D_2$ lines, see e.g.~\cite{deutsch1998,mandel2003,gadway2010}. However such a scheme is efficient only if the two considered state have different magnetic susceptibilities~\cite{grimm2000}. Since we use in this work the  ``clock states" $  |1\rangle = |F=2,m_F=1\rangle $ and $ |2\rangle = |F=1,m_F=-1\rangle $, another method has to be implemented.


\begin{figure*} 
		\centering
		\includegraphics[width=0.8\textwidth]{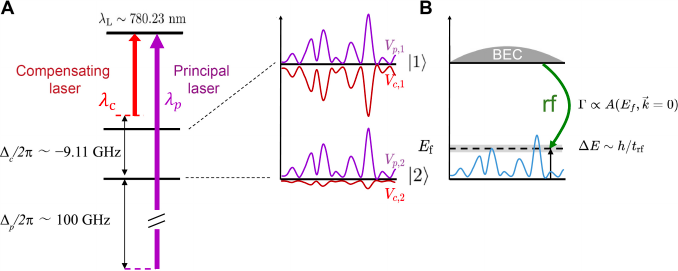} 
		
		\caption{\textbf{Energy-resolved scheme based on state-dependent disorder and weak rf transfer.} 
\textbf{(A)}~Laser detuning configuration of the bichromatic speckle field : a blue-detuned -- principal -- beam  ($\Delta _p/2\pi\approx$\SI{100}{GHz} with respect to the D2 line of $^{87}$Rb atoms) and a red-detuned -- compensating -- beam ($\Delta _c/2\pi\approx$\SI{-9.11}{GHz}) are used to shine two almost identical laser speckle fields on the atoms. The intensities are adjusted to cancel disorder in state $|1\rangle$ (the ``free" state) but not in state $|2\rangle$ (the ``disorder-sensitive" state).
\textbf{(B)}~A small fraction of atoms initially prepared in state $|1\rangle$ are transferred to  state $|2\rangle $  using a weak rf coupling field for a duration $t_{\rm rf}$. This allows us to selectively populate energy states in the disordered potential, with the targeted energy $E_{\rm f}= \hbar \delta_{\rm f}$ and a Fourier-limited energy resolution $\Delta E \sim h/t_{\rm rf} $. In this weak coupling regime, the transfer rate $\Gamma$, which can be estimated using the Fermi golden rule, is proportional to the spectral function $A (E, \mathbf{k=0})$ (see Fig.~\ref{fig:spectralfunction}).}
		\label{fig:Bichromatic} 
	\end{figure*}

\subsubsection{Principle of the bichromatic laser speckle field}

The state-dependent disorder is created using a bichromatic laser speckle configuration where two almost identical laser speckle fields, referred to as the principal and the compensating one, are generated at slightly different wavelengths. As illustrated in Fig.~\ref{fig:Bichromatic}A , the principal and the compensating lasers are respectively detuned by $\Delta _p/2\pi\approx$\SI{100}{GHz} and $\Delta _c/2\pi\approx$\SI{-9,1}{GHz} from the $D_2$ line of $^{87}\mathrm{Rb}$ ($\lambda_L$=\SI{780.23}{nm}). In this configuration, the principal laser yields repulsive speckle potentials, $V_{\mathrm{p},1}(\mathbf{r})$ and $V_{\mathrm{p},2} (\mathbf{r})$, of similar amplitudes (except for the coupling strength of each transition) for both states $ |1\rangle  = |F=2,m_F=1\rangle $ and $ |2\rangle = |F=1,m_F=-1\rangle $. In contrast, the compensating laser yields attractive speckle potentials, $V_{\mathrm{c},1} (\mathbf{r}) $ and $V_{\mathrm{c},2} (\mathbf{r})$, but with an absolute amplitude much stronger in state $|1\rangle$ than in state $|2\rangle$.  The laser intensities are then adjusted to achieve (i) an almost perfect cancellation in the ``free" state $|1\rangle $ and (ii) the targeted disorder potential $V(\mathbf{r})$ (of amplitude $\langle V \rangle =\VR$, see section~\ref{Sec:Speckle}, for the ``disorder-sensitive" state $|2\rangle $:
\begin{eqnarray}
V_1(\mathbf{r}) &= & V_{\mathrm{p},1} (\mathbf{r}) + V_{\mathrm{c},1}  (\mathbf{r})\sim 0   \label{Eq:statedependantZero}\\ 
 V_2(\mathbf{r}) &=&V_{\mathrm{p},2} (\mathbf{r}) + V_{\mathrm{c},2} (\mathbf{r})=V(\mathbf{r}) \; .\label{Eq:statedependantVr}
\end{eqnarray}
As an example, the amplitudes  $\langle V_{\mathrm{p},i} \rangle $ and $\langle V_{\mathrm{c},i}\rangle $ corresponding to the case $\VR/h=\SI{416}{Hz}$ are given in table~\ref{tab:disorder_comparison}. As discussed in section~\ref{Sec:SpectralFunction}, the disorder amplitude $\VR$ is precisely calibrated in the experiment using spectral functions measurements, yielding around 5$\%$ uncertainty.

The scheme is similar to the one demonstrated in Ref.~\cite{lecoutre2022bichromatic}, except for the inversion of the ``free" and ``disorder-sensitive" states. This choice is motivated by the presence of spin-exchange collisions in state $|1\rangle$, which lead to two-body losses and heating~\cite{Egorov2013}, thereby preventing the use of this state to probe transport properties in a disordered potential over long timescales. In contrast, spin-exchange collisions do not occur for state $|2\rangle$.

Note that the laser detunings are chosen as a compromise between the photon-scattering-limited lifetime in state $|2\rangle$ (see section~\ref{Sec:Lifetime}) and the fundamental spatial decorrelation of the principal and compensating laser speckle fields. Indeed, a key aspect of the scheme is the cancellation of disorder in state $|1\rangle$, which cannot be perfect because the two speckle fields are generated at different wavelengths. Nevertheless, as shown in Ref.~\cite{lecoutre2022bichromatic}, this decorrelation can be neglected in our configuration, since the relative wavelength difference $\delta \lambda/\lambda \sim 2.5\times 10^{-4}$ is sufficiently small. More quantitatively, we estimate that state $|1\rangle$ experiences a residual disorder with a rms amplitude $\sigma_{V_1}$ on the order of 0.02~$\VR$. For the specific case $V_\mathrm{R}/h=\SI{416}{Hz}$, this corresponds to about $\sigma_{V_1} \sim \SI{10}{  \hertz}$.

\subsubsection{Life time optimization  \label{Sec:Lifetime}}

The detunings of the principal and compensating lasers are chosen to optimize the lifetime $\Gamma_2^{-1}$ in state $|2\rangle$, subject to two important constraints: (i) the lifetime of state $|1\rangle$ must be sufficiently long to avoid any significant perturbation of the BEC state during the rf transfer, i.e. $\Gamma_1^{-1} > t_\mathrm{rf}$, and (ii) the relative wavelength difference is fixed at $\delta \lambda/\lambda \sim 2.5 \times10^{-4}$ so that the spatial decorrelation between the two laser speckle fields can be neglected (see above). For the specific case of $V_\mathrm{R}/h=\SI{416}{\hertz}$, this yields $\Gamma_1^{-1}=\SI{125}{\milli \second}$ and  $\Gamma_2^{-1}= \SI{1.17}{\second}$ (see table~\ref{tab:disorder_comparison}). Such a lifetime in state $|2\rangle$ is not long enough to efficiently probe transport properties -- localised versus diffusive -- in the disordered potential. It is mainly limited by photon scattering from the compensating laser, which is closer to resonance. 

\begin{figure} 
		\centering
		\includegraphics[scale=0.75]{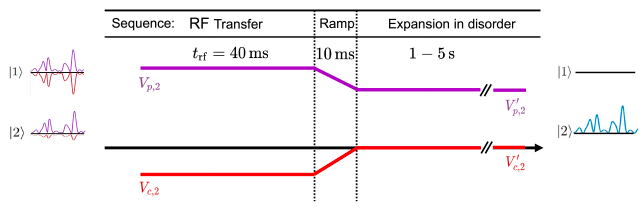} 
		\caption{
\textbf{Improvement of the atomic lifetime during expansion in the disorder.} During the rf transfer stage, the principal and compensating beam intensities are tuned to realize a state dependent-disorder, as described in Fig.~\ref{fig:Bichromatic}. The lifetime in state $|2\rangle $ is nevertheless limited by the presence of the compensating beam, whose presence becomes useless during the expansion in disorder. Once the rf transfer is over, we then ramp down the compensating disordered potential to $V'_{c,2}$=0 over \SI{10}{ms}, while ramping down the principal beam intensity such that the total disorder amplitude in state $|2\rangle $ remains constant to $\VR$ (see Eq.~\ref{Eq:statedependantVr}). This allows us to increase significantly the lifetime, up to several seconds (see Tab. \ref{tab:disorder_comparison}).}
		\label{fig:Ramp} 
	\end{figure}

The compensating laser is, however,  no longer required once the rf transfer is switched off at $t=0$. The lifetime in state  $|2\rangle$ is then further increased by implementing a ramp sequence immediately after the rf transfer, as shown in Fig.~\ref{fig:Ramp}. During this stage, the compensation potential $V_{c,2}$ is linearly ramped down to $V'_{c,2} = 0$ over $\SI{10}{\milli \second}$, while  the principal disordered potential is simultaneously adjusted from $V_{p,2}$ to $V'_{p,2}=\VR$. These concomitant intensity ramps preserve the total disorder amplitude experienced by state $|2\rangle$ throughout the sequence and have a negligible impact on the subsequent dynamics. For the specific case $\VR/h = \SI{416}{\hertz}$, we estimate an improved lifetime of $\Gamma_2^{-1}=\SI{7.2}{\second}$. Experimentally, we measure a lifetime of $\Gamma_2^{-1}=\SI{5(0.5)}{\second}$ by repeating the same sequence while keeping the optical trap on to freeze the dynamics. The slight discrepancy with the calculated value is attributed to the finite vacuum lifetime, measured around~$\SI{20}{\second}$.

\begin{table}[htbp]
\centering
\label{tab:potentials}
\renewcommand{\arraystretch}{1.3}
\resizebox{\columnwidth}{!}{
    \setlength{\tabcolsep}{2pt} 
    \begin{tabularx}{1.05\columnwidth}{@{} l *{8}{c} @{}} 
    \hline\hline
    & \multicolumn{4}{c}{\textbf{Potential on state $|1\rangle$}} 
    & \multicolumn{4}{c}{\textbf{Potential on state $|2\rangle$}} \\
    \cline{2-5} \cline{6-9}
    \textbf{Sequence} 
    & $V_{p,1}/h$ & $V_{c,1}/h$ & $V_1/h$ & $\Gamma_1^{-1}$ 
    & $V_{p,2}/h$ & $V_{c,2}/h$ & $V_2/h$ & $\Gamma_2^{-1}$ \\
    \hline
    rf transfer & 535 & -535 & 0 & 125\,ms 
                & 569 & -153 & 416 & 1.17\,s \\
    Expansion & \color{gray}- & \color{gray}- & \color{gray}- & \color{gray}- 
                & 416 & 0 & \textbf{416} & \textbf{7.2\,s} \\
    \hline\hline
    \end{tabularx}
}
\caption{\textbf{State-dependent disordered potentials and scattering lifetimes.}
Comparison of the disordered potentials experienced by states $|1\rangle$ and $|2\rangle$ for a disorder amplitude of $\VR/h=\SI{416}{Hz}$. The parameters are calculated during the rf transfer sequence (first row) and during the expansion of atoms in state $|2\rangle$ in the disorder  (second row), i.e., after the disorder ramp sequence shown in Fig.~\ref{fig:Ramp}.
$V_{p,i}/h$ and $V_{c,i}/h$ denote the contributions of the principal and compensating laser beams to the average disordered potential acting on state $|i\rangle$, while $V_{i}/h$ is the total averaged potential. For simplicity, the $\langle ... \rangle$ averaging notation is omitted.
$\Gamma_i^{-1}$ is the photon-scattering-limited lifetime of state $i$.
The parameters for state $|1\rangle$ during the expansion are not shown, the atoms being in state $|2\rangle$. } 
\label{tab:disorder_comparison}
\end{table}

\subsubsection{Laser speckle properties \label{Sec:Speckle}}

{\it Geometry --- }
The laser speckle fields are generated in the same manner for both the principal and the compensating beams. To ensure that both fields illuminate the diffuser identically, the two beams are injected into the same single-mode optical fiber before being expanded and collimated, resulting in identical Gaussian beams with a $1/e^2$ radius of \SI{9(1)}{mm} propagating along the $x$-axis. The beams are then incident on a diffuser, which is truncated by a circular diaphragm with a diameter of \SI{20.3(1)}{mm}, and subsequently focused onto the atoms located \SI{15.2(5)}{mm} away. This geometry corresponds to a high numerical aperture $\mathrm{NA}=0.55(2)$.

At the position of the atoms, each speckle field exhibits a Gaussian average intensity envelope with a $1/e^2$ radius of \SI{1.47}{mm}. This is much larger than the spatial extent of the atomic cloud, ensuring negligible intensity variation of the disorder across the sample. 
\\

{\it Potential distribution $P(V)$ --- }
 Laser speckle fields are characterized by an exponential intensity distribution~\cite{goodman2007speckle}. For a ``repulsive" (positive) speckle, as used in the experiment, the intensity distribution reads $P(V)=1/\langle V \rangle \exp(-V/ \langle V \rangle ) \Theta(V)$, where $\Theta$ denotes the Heaviside step function. This distribution is characterized by a rms value, referred to as the disorder amplitude $\VR$, which is equal to its mean value, i.e. $\VR=\langle V \rangle$. 
\\

{\it Spatial correlation lengths ---}
The disorder spatial correlation is determined from the two-point intensity correlation function $\langle I(\mathbf{r}) I(\mathbf{r}  + \mathbf{ \delta r}  ) \rangle$, which  has been measured experimentally, see e.g., Ref.~\cite{Volchkov2018}. Following the convention adopted in Ref.~\cite{Ec_Delande2017}, the measured HWHM values are converted to the correlation lengths $\sigma_{\perp} = \SI{0.30}{\micro \meter}$  (for the transverse correlation function in the $y-z$ plane) and $\sigma_{\parallel}= \SI{1.45}{\micro \meter}$ (for the longitudinal correlation function along the $x$-axis). The characteristic energy scale,  called the correlation energy, is then given by \begin{equation}
E_\sigma = \frac{\hbar^2}{m\bar{\sigma}^2} \simeq h \times 441~\mathrm{Hz} \;.
\end{equation}
Here $\bar{\sigma}=(\sigma_{\perp}^2\sigma_{\parallel})^{1/3}= \SI{0.51}{\micro \meter}$  is the geometric mean of the correlation lengths in the three directions.

\subsection{Energy-resolved state preparation}

 \subsubsection{Principle of the transfer scheme \label{sec:rftransfer}}

The energy-resolved scheme relies on driving an rf transition from the ``free" disorder state $|1\rangle= |F=2,m_F=1\rangle $ to the disorder-sensitive state $|2\rangle= |F=1,m_F=-1\rangle $ (see Sec.~\ref{Sec:StateDependent}). The transition is implemented via a two-photon rf-process involving the intermediate state $|0\rangle=|F=2,m_F=0\rangle$. As detailed in Ref.~~\cite{Volchkov2018}, a first microwave field of frequency $\omega_1/2\pi \sim \SI{6.83}{\giga \hertz}$ couples the $|1\rangle\rightarrow|0\rangle$ with a detuning  $\delta_0/ 2\pi\sim\SI{0.5}{\mega \hertz}$.  A second rf field of frequency $\omega_2 /2\pi  \sim \SI{2.8}{\mega \hertz} $ then drives the  $|0\rangle\rightarrow|2\rangle$ transition. The couplings are sufficiently weak that the system reduces to an effective single rf transition from  $|1\rangle\rightarrow|2\rangle$, driven at the total frequency $\omega_{\rm rf}=\omega_1+\omega_2$, with the effective Rabi frequency $\Omega_{\rm rf}=\Omega_1\Omega_2/2\delta_{0}$ ($\Omega_{1,2}$ denote the Rabi frequencies of the individual fields). 

By tuning the total rf frequency $\omega_{\rm rf}$, the atoms are selectively populated into the eigenstates of the disordered potential at a given energy $E_{\rm f}=\hbar \delta_{\rm rf}$, where $\delta_{\rm rf}= \Delta_{\rm HFS} - \omega_{\rm rf}$ is the detuning from the bare resonant frequency (that is in the absence of disorder) corresponding to the hyperfine splitting between the respective internal energies $\Delta_{\rm HFS}/h \sim \SI{6.8}{GHz}$ (see Fig.~\ref{fig:Bichromatic}B).

The rf transfer is implemented in the presence of a bias magnetic field $B^\star\approx$ \SI{3.23}{G}. At such a ``magic" magnetic field, both states $|1\rangle$ and $|2\rangle$ experience identical magnetic susceptibilities. This property is crucial for suppressing magnetic noise in the  $|1\rangle\rightarrow|2\rangle$ transition, enabling a frequency precision at the level of a few Hertz. Moreover, as discussed in Sec.~\ref{Sec:Levitation}, it allows both states to be simultaneously levitated against gravity.

Finally, as discussed in detail in~\cite{Volchkov2018}, it should be noted that the transfer scheme is almost immune to interactions within the mean-field approximation. This is due to a specific property of $^{87}$Rb atoms, where the intra- and inter-state $s$-wave scattering lengths are nearly equal for the states $|1\rangle$ and $|2\rangle$, with $a_{11} \approx a_{22}  \approx a_{12} \approx 100 \; a_0$, where $a_0$ is the Bohr's radius. First, this near-equality ensures that that the total energy changes very little when an atom is transferred from $|1\rangle$ to $|2\rangle$. Consequently, interactions have a negligible effect on the resonance condition $E_{\rm f}= \hbar \delta_{\rm rf}$. Accordingly, in the absence of disorder, the system exhibits high-contrast Rabi oscillations with a coherence time extending to several hundred milliseconds. Second, in the Thomas-Fermi regime, the mean-field interaction precisely compensates the harmonic trapping potential. Since the atoms in state $|1\rangle$  and $|2\rangle$ experience the same mean-field interaction, the effect of the harmonic potential is effectively canceled. This allows $|1\rangle$ to be treated as a ``free" state, and state $|2\rangle$ as ``disorder-sensitive", as shown in Fig.~\ref{fig:Bichromatic}B.

\subsubsection{Measurement of the spectral function $A (E, \mathbf{k=0})$ \label{Sec:SpectralFunction}}
 
 \begin{figure} 
		\centering
		\includegraphics[scale=0.52]{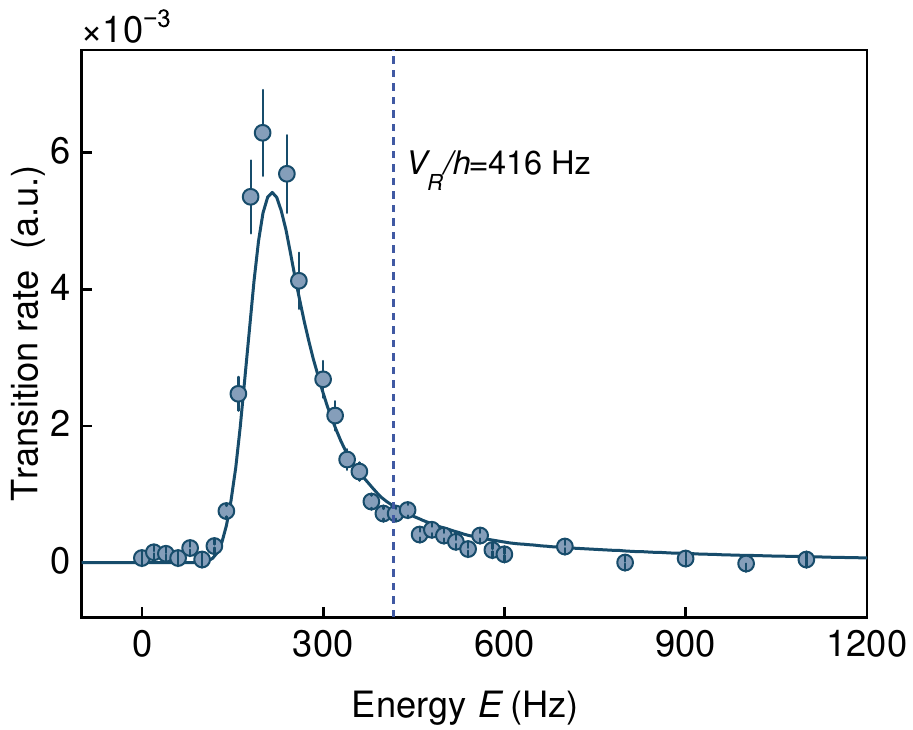} 
		\caption{\textbf{Spectral function $A (E, \mathbf{k=0})$ for the disorder amplitude  $\VR/h=$\SI{416}{Hz}.}
			Blue circles show the normalized rf transition rate $\Gamma$ extracted from the number of atoms transferred from $\lvert 1\rangle$ to $\lvert 2\rangle$. The solid line represents the independently numerically calculated spectral function for our 3D laser speckle configuration, with no free parameters. It is convolved with the energy filter function $F$ associated with the Kaiser-shape pulse window used in the experiment, see Eq.~\ref{Eq:KaiserFilter}.  Both data sets (experiment and numerics) are normalized to unit integral for proper comparison. In practice, this comparison allows for a calibration of the disorder amplitude with a $\sim 5\%$ accuracy. 	}
		\label{fig:spectralfunction} 
	\end{figure}

 In the presence of disorder, the spectrum in state $|2\rangle$ becomes quasi-continuous. As shown in Ref.~\cite{Volchkov2018}, we can then model our rf transfer scheme as the irreversible coupling from a discrete initial state -- the BEC in state $|1\rangle$ -- to a continuum. In the weak coupling regime, where the effective Rabi coupling $\Omega_{\rm rf}$ is much smaller than the typical width of the continuum, the transition rate $\Gamma$ is given by the Fermi's Golden Rule:
 \begin{eqnarray}
       \Gamma (\delta_{\rm rf } )  &\propto& \overline{ \langle \psi_{\rm BEC} | \psi_{E_{\rm f}  }\rangle} ^2  \rho(E_{\rm f}) \, \ast \, F(E) \nonumber  \\
&\sim& \overline{ \langle \mathbf{k=0}  | \psi_{E_{\rm f}  }\rangle} ^2  \rho(E_{\rm f}) \, \ast \, F(E) \nonumber \\
 &=& A (E_{\rm f} , \mathbf{k=0}) \, \ast \, F(E) \;  .\label{Eq:SpectralFunction}
 \end{eqnarray}
 Here, $\psi_{E_{\rm f}}$ denotes the eigenstate wavefunctions at energy $E_{\rm f}$, $\overline{\cdots}$ refers to averaging over disorder, and the BEC wavefunction is approximated as uniform, $| \psi_{\rm BEC} \rangle \sim | \mathbf{k=0} \rangle$. This approximation is justified by the large spatial extent of the BEC compared to the speckle correlation length. Within this approximation, Eq.~\eqref{Eq:SpectralFunction} shows that the transfer rate is directly proportional to the spectral function $A (E_{\rm f}, \mathbf{k=0})$, up to a convolution with the energy filter function $F(E)$ (the squared  Fourier transform of the rf pulse time envelope $f(t)$; see Sec.~\ref{sec:EnergyDistribution}). The spectral function corresponds to the probability for a state of energy $E_{\rm f} $ to have the momentum $\mathbf{k=0}$. This has been experimentally demonstrated in Ref.~\cite{Volchkov2018}, through direct comparison between measured transfer rates and independently calculated spectral functions, showing excellent agreement over a wide range of disorder strengths (from $\eta=\VR/E_\sigma= 0.1$ to 10). 
 
 Fig.~\ref{fig:spectralfunction}A shows such a comparison for the case $\VR/h=\SI{416}{\hertz}$ and the specific scheme used in this work. The transfer rate is measured directly by counting the transferred atoms in state $|2\rangle$ for a weak rf field  and a short coupling time such that $\Gamma t_{\rm rf}  \ll 1$. In this regime, the transferred atom number grows indeed linearly with time, with $N_2(t)=N_1(0) \Gamma t_{\rm rf}  $. Independently, the spectral function is  computed numerically for an ideal plane wave $|\mathbf{k} = \mathbf{0} \rangle$ launched in a 3D speckle potential having the same spatial correlation properties as in the experiment (same method as in Ref.~\cite{Volchkov2018}). The agreement is here again very good. 
 
Finally, noting that the shape of the spectral function is highly sensitive to the disorder amplitude, we use this comparison in practice to calibrate $\VR$, with an uncertainty of 5$\%$. This provides a significantly more accurate calibration than standard photometric measurements, which are typically limited to around 20$\%$ uncertainty. 

\subsubsection{Energy distribution \label{sec:EnergyDistribution}}

{\it Fourier-limited energy distribution ---} 
The key feature of the rf transfer scheme is the preparation of energy states with a very narrow energy distribution. When operating in the weak-coupling regime and at short times $\Gamma t_{\rm rf}  \ll 1 $ (i.e., transferring only a small fraction of the atoms, as is the case throughout this work), the energy spread of the transferred atoms is Fourier-limited to $\Delta E \sim  h / t_{\rm rf}$. More precisely, the energy distribution $\mathcal{D}(E;E_{\rm f})$, associated with the targeted energy $E_{\rm f}= \hbar  \delta_{\rm rf}$, is given by: 
\begin{equation}
     \mathcal{D}(E;E_{\rm f}) =  A (E, \mathbf{k=0}) \, F(E - E_{\rm f}) \; .
\label{eq:supEDistrib}
\end{equation}
Here  $A (E, \mathbf{k=0}) $ is the spectral function associated with $\mathbf{k=0}$, as discussed in the previous section. $F(E)$ is here again the ``energy filter" function corresponding to the pulse shape function $f(t)$ (see next section). When the spectral function is smooth, the energy distribution $\mathcal{D}(E;E_{\rm f})$ is essentially determined by the filter function $F(E)$, such that the distribution is centered around $E_{\rm f}$ with a typical width $\sim h / t_{\rm rf}$. Nevertheless, as we show below, the peaked structure of the spectral function---see Fig.~\ref{fig:spectralfunction}A---has a noticeable effect at low transfer energies $E_{\rm f}$, leading to a slight energy shift. 
\\

{\it Kaiser-shape rf pulse --- } 
In order to suppress side lobes, we shape the rf pulse with a Kaiser window function given by:
\begin{equation}
	f(t)=\frac{I_0\!\left[2 \pi \sqrt{1-\left({2t}/{t_{\mathrm{rf}}}-1\right)^2}\right]}{I_0(2\pi)}, 
	\qquad 0<t< t_{\mathrm{rf}} \; ,
\end{equation}
where $I_0(x)$ is the modified Bessel function of zeroth order. Its Fourier transform is 
 \begin{equation}
    F(E)
    = C \: \mathrm{sinc}^{2}\!\left(
        \sqrt{
            \left(\frac{\pi t_\mathrm{rf}E}{h}\right)^2
            - (2 \pi)^2
        }
    \right) \: , \label{Eq:KaiserFilter}
\end{equation}
with the definition $\mathrm{sinc} (x) = \sin(x)/x$. Here $C$ is a normalization constant ensuring the proper normalization of the energy distribution. For $t_\mathrm{rf} = 40$~ms, the rms width of $F$ is $\Delta E /h \simeq \SI{13.5}{\hertz}$. 
\\

\begin{figure} 
		\centering
		\includegraphics[scale=0.7]{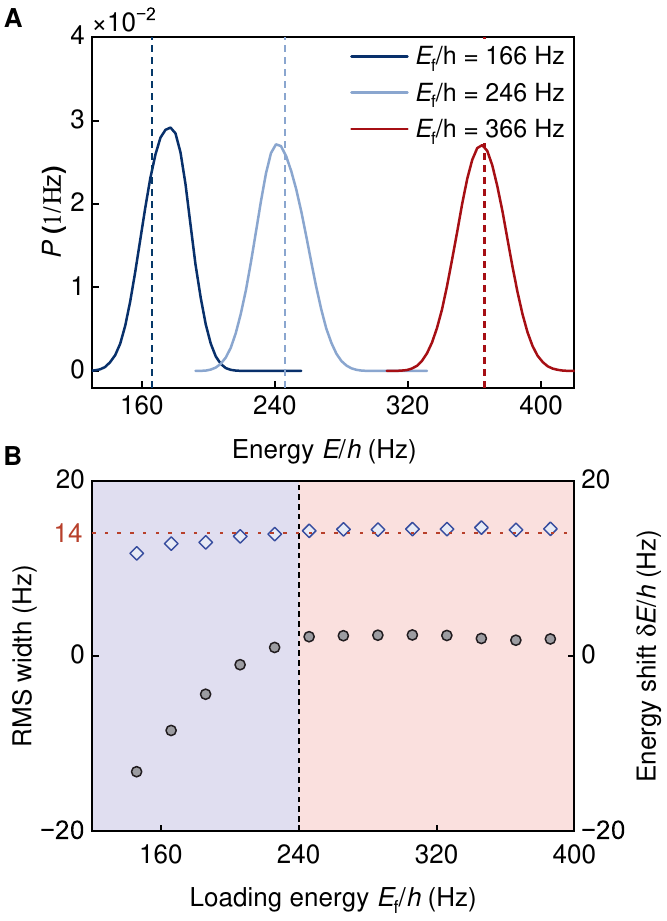} 
		\caption{\textbf{Estimated energy distribution of the atoms in the disorder potential of amplitude $\VR/h=\SI{416}{\hertz}$.}
        \textbf{(A)}~Energy distribution $\mathcal{D}(E;\Ef)$ calculated following Eqs.~\eqref{eq:supEDistrib} and \eqref{Eq:KaiserFilter} for the targeted energies $E_{\rm f}/h$=166, 246 and 366~Hz (vertical dashed lines).
        \textbf{(B)} Light blue diamonds denote the rms width $\Delta E$ of the calculated energy distribution $\mathcal{D}(E;\Ef)$. It is approximately constant across the entire energy range, with values close to \SI{14}{Hz} (horizontal red dotted line), in agreement with the rms width of energy filter function $F$. The grey circles show the energy shift $\delta E=\Ef-\bar{E}$ between the targeted energy $\Ef$ and the effective mean energy $\bar{E}$ of the energy distribution. At low $\Ef$~(shaded blue region), the sharp variation of the spectral function leads to a negative energy shift. It gradually decreases as $\Ef$ approaches the expected mobility edge (vertical dashed line), and then becomes negligible (well below the width $\Delta E$) at higher energies.}
\label{fig:EnergyDistribution}
	\end{figure}

{\it Energy width and estimated residual shift --- }
Fig.~\ref{fig:EnergyDistribution}A shows the energy distribution $\mathcal{D}(E;\Ef)$  for several targeted energies $E_{\rm f}$ at a disorder amplitude $\VR/h=\SI{416}{\hertz}$. The distribution is estimated using Eqs.~\eqref{eq:supEDistrib} and \eqref{Eq:KaiserFilter}, with the spectral function $A (E, \mathbf{k=0})$---fixed for a given disorder amplitude---numerically computed as described in Sec.~\ref{Sec:SpectralFunction}. 

Two important aspects should be noted. First, the rms energy width $\Delta E \sim  \SI{13}{\hertz}$ of the distribution $\mathcal{D}(E;\Ef)$ is nearly constant across the range of  energies $E_{\rm f}$, and it is consistently very close to the rms width of the filter function $F(E)$. Second, the mean energy $\bar{E}$ of the energy distribution does not exactly coincide with the targeted energy $\Ef$ when transferring the atoms at low or high energy. This effect is due to the non-negligible variation of the spectral function over the typical energy width $\Delta E$. It is particularly pronounced at low energy, where the spectral function exhibits a steep slope due to the low density of states in this regime.

To be quantitative, the estimated energy shift $\delta E = \Ef - \bar{E}$ is shown in Fig.~\ref{fig:EnergyDistribution}B for the range of transferred energies $E_{\rm f}$ considered in this work. This shift is only significant at very low energy, where it exceeds the rms width.  We have verified, however, that taking this shift into account does not affect the analysis of the data presented in the main text, and in particular has no impact on the determination of the mobility edge. This is  because the reported observations are largely insensitive to the energy in the low-energy ---localised--- regime (see, e.g., Fig.~\ref{fig:ME_measure} of the main text). Therefore, we did not take into account this shift in the analysis.

\subsection{Magnetic levitation \label{Sec:Levitation}}

Throughout the experiment, we use a magnetic levitation to suspend the atoms against gravity, regardless the internal states $| 1 \rangle$ and $| 2 \rangle$. To do so, we combine a magnetic field gradient $b^\prime $ along the $z$-axis ---created by a pair of vertical coils in an anti-Helmholtz configuration--- , together with a homogeneous vertical bias field ---created by another pair of vertical coils in a Helmholtz configuration. The bias field is set to the ``magic" magnetic field $B^\star=\SI{3.23}{G}$, a value for which the states $| 1 \rangle$ and $| 2 \rangle$ have exactly the same magnetic susceptibility. At low magnetic field, the susceptibility is very close to the one given by the linear Zeeman effect, that is given by $ m_F g_F \mu_{\rm B}$, where $g_F$ is the Land factor and $ \mu_{\rm B}$ the Bohr's magneton.  By setting the magnetic field gradient to $b'\sim m g/ m_F g_F \mu_{\rm B} \sim  \SI{30.5} \, \mathrm{G/cm}$, we thus compensate precisely the gravity for both states at the same time. 

 The vertical magnetic gradient is, however, unavoidably associated with a horizontal gradient $b'/2$ along the $x$ and $y$ axis. The latter introduces a spatial inhomogeneity of the magnetic field as one moves away from the vertical axis of symmetry, leading to a residual harmonic confinement in the horizontal plane of frequency $\omega_\perp$. The atoms being locally sensitive to the norm of the magnetic field, a standard calculation gives~\cite{bernard2010these}:
\begin{equation}
		\omega_\perp =  \sqrt{\frac{m g^2}{ 4 m_F g_F \mu_{\rm B} B^\star}} \approx 2\pi\times 7~\mathrm{Hz}.
\end{equation}
Note that we recover here the usual $1/\sqrt{B^\star}$ dependence for the transverse frequency in a magnetic levitation scheme~\cite{leanhardt2003cooling,sackett2006limits}. Along the $z$-direction, the confinement---which would normally arise from any curvature $b"$ of the vertical magnetic component---is effectively canceled by the (anti-) Helmholtz coil configuration, yielding $\omega_z \simeq 0$.




\subsection{Inital state preparation }

\subsubsection{BEC parameters in state $|1\rangle$ \label{Sec:BEC}}

The first step of the experimental sequence is the creation of a BEC of $\sim1.9\times 10^5$ $^{87}\mathrm{Rb}$ atoms in the hyperfine state $|1\rangle$. The BEC is confined in a nearly isotropic optical trap, with trapping frequencies $(\omega_x, \omega_y, \omega_z) \approx 2\pi \times (35, 25, 11)$ Hz. The temperature is around $T \sim 7\,\mathrm{nK}$, yielding a condensate fraction typically around 75$\%$ and a chemical potential $\mu_{\rm in}/h\approx $ \SI{350}{Hz}.  These parameters are determined from the time of flight expansion of the atomic cloud in the presence of the residual weak transverse confinement associated with the magnetic levitation (see Sec.~\ref{Sec:Levitation} above), and a comparison with the predictions from the scaling laws for the condensed fraction~\cite{castin1998low}. Accordingly, we estimate Thomas-Fermi radii of $R_{\rm TF} \approx \SIlist{8; 12; 26}{\micro \meter}$ along the respective axes, this estimation being more reliable than the direct in-situ fluorescence imaging of the BEC. The latter can indeed be impaired by multi-photon scattering processes and residual distortions of the profiles along the $y$ axis due to the radiation pressure experienced by the atoms during the imaging pulse of \SI{20}{\micro \second} (see Fig.~\ref{fig:twoDexpansion}).

\subsubsection{Initial profiles in state $|2\rangle$ at $t=0$ \label{Sec:InitialProfile}}

\begin{figure} 
			\centering
			\includegraphics[scale=0.38]{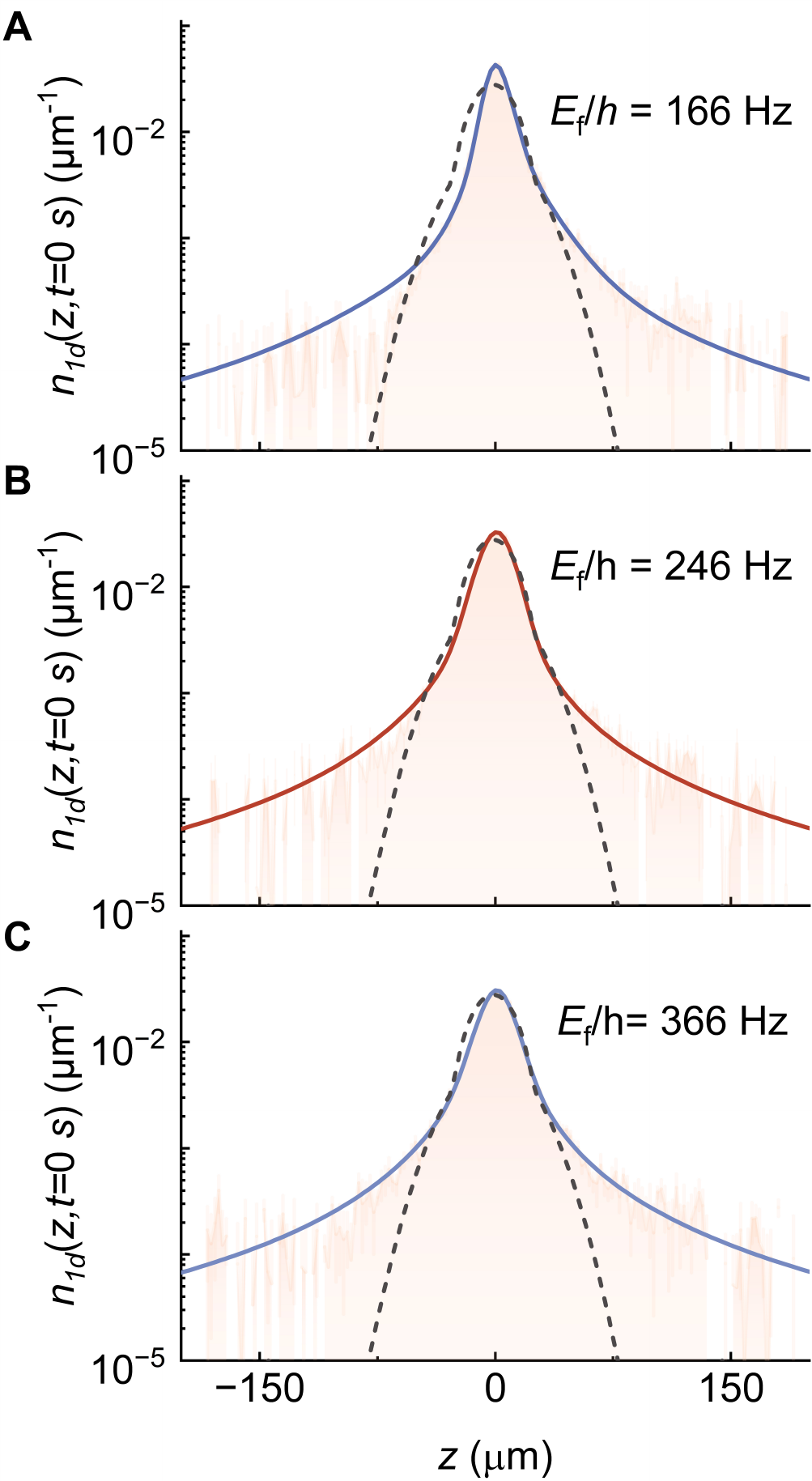} 
			\caption{\textbf{Initial profiles for the atoms in state $|2\rangle$ at $t=0$.} Orange: Initial integrated 1D density profiles along the $z$ axis, $n_{\rm 1d}(z, t=0,\mathrm{s})$, for three loading energies: $\Ef/h=$166 (A), 246 (B) and $\SI{366}{\hertz}$ (C). The profiles are shown on a semi-log scale, as in Fig.~\ref{fig:Transition} of the main text. The dashed lines correspond to the time-of-flight estimation of BEC shape, including the thermal component, in state $|1\rangle$ before rf transfer (see Sec.~\ref{Sec:BEC}). Both profiles have been normalized to unit integral  for a proper comparison. A slight variation is observed on the initial profile in state $|2\rangle $ (after rf transfer), the profile being slightly narrower than the BEC at low energy (see text). The solid curves represent the pseudo-Voigt fits, used as input for the self-consistent theory calculations (see Sec.~\ref{Sec:SCT}).}
			\label{fig:InitialZprofile} 
		\end{figure}

Following the BEC preparation, a small fraction, with typically $N \approx 10^4$ atoms, is transferred to the target state $|2\rangle$ using the rf pulse (see Sec.~\ref{sec:rftransfer}). The atomic density being much smaller, we directly image the atomic profile at $t=0$ ---just after the end of the rf pulse--- using in-situ fluorescence imaging. Fig.~\ref{fig:InitialZprofile} shows the corresponding integrated 1D profile $n_{\rm 1D} (z,t=0)$ along the $z$-axis for the energies $\Ef/h=$166, 246 and $\SI{366}{\hertz}$. Note that the first and the third one correspond to the initial profiles shown on Fig.~\ref{fig:Transition}B of the main text. 

Two important aspects must be emphasized. First, although the initial profiles are close to the BEC shape (as inferred from time-of-flight measurements, see above), slight variations are nevertheless observed depending on the loading energy $\Ef$. At low energy, the initial cloud size is slightly smaller than the BEC size, while it converges to the same size as the energy increases. This behaviour is also clearly visible in Figs.~\ref{fig:Dynamics} of the main text. In particular, Fig.~\ref{fig:Dynamics}B shows a slight increase of the experimental cloud size $\sigma_{\rm exp}$ with $\Ef$, saturating for $\Ef\gtrsim \SI{266}{\hertz}$. This behavior can be attributed to both (i) the decrease of the density of states at low energy and (ii) the smaller spatial extent of localized states at low energy. As a result, the rf coupling between states $|1\rangle$ and $|2\rangle$ may not be fully spatially homogeneous across the BEC, leading to  a slightly reduced size of the transferred atomic cloud. 

Second, we observe the presence of weak tails in the initial profiles at $t=0$, extending beyond the BEC, including the thermal component, prepared in state $| 1 \rangle$. We attribute this effect to the dynamics that may occur in state $|2\rangle$ during the rf coupling process, as well as to possible short-time atomic interaction effects. A precise numerical and experimental investigation of the coupling dynamics is left for future work to clarify this aspect. Importantly, this does not affect the analysis presented in this paper, as the directly measured $n_{\rm 1D} (z,t=0)$ distribution is used as the initial profile. 

\subsubsection{Estimation of the interaction energy in state $|2 \rangle$ at $t=0$}

We estimate the interaction energy $E_{\rm int}$ based on the initial density in state $|2\rangle$ after rf transfer, assuming that the 3D profile is identical to that of the BEC in state $|1\rangle$ (i.e. with the Thomas-Fermi radii $R_{\rm TF} \approx \SIlist{8; 12; 26}{\micro \meter}$). While this assumption may slightly overestimate the interaction energy -- as discussed above, there is a slight difference the transferred cloud has slightly expand on the wings, see Fig.~\ref{fig:InitialZprofile}--, it allows us to account for the cloud size along the $x$ axis, which is not accessible from our imaging setup. This yields 
\begin{equation}
E_{\rm int} = \frac{g_{\rm eff} }{2} \int n^2(\mathbf{r}) d\mathbf{r} \; \approx h \times \SI{5}{\hertz} \; ,
\end{equation} 
where $g_{\rm eff} = 4\pi\hbar^2 a/m$, with $a \approx 100\,a_0$ the $s$-wave scattering length of $^{87}\mathrm{Rb}$. This interaction energy is much smaller than the other relevant energy scales and further decreases over time due to the combined effects of cloud expansion and uniform atom losses (with lifetime $\tau\sim\SI{5}{\second}$, see Sec.~\ref{Sec:Lifetime}). Consequently, it is unlikely to play a significant role during the expansion. It is, however, not excluded that the local interaction energy may exceed this estimate, particularly when atoms are loaded into localized states, which are likely to have sharply peaked densities. In such cases, these interactions could induce dynamics during the rf coupling process, potentially contributing to the formation of the wings discussed above, and might slightly affect the energy distribution. A detailed analysis of these coupling dynamics is left for future work.

\subsection{Expansion of the atomic cloud in the disorder}

\subsubsection{Confinement in the horizontal $x-y$ plane \label{sec:confinement}} 

 \begin{figure} 
		\centering
		\includegraphics[scale=1.55]{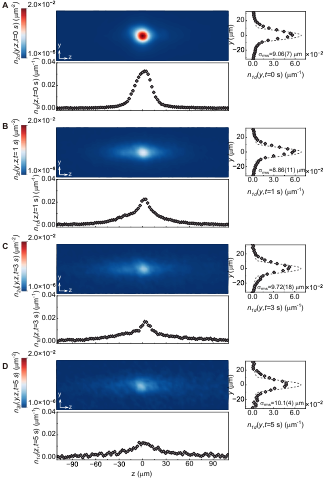} 
		\caption{
		\textbf{Expansion of the atomic cloud at $E_{\rm f}/h=\SI{246}{\hertz}$ for the disorder amplitude $\VR/h=\SI{416}{\hertz}$.} (A–D) Two-dimensional density profiles recorded after  (A)~\SI{0}{s}, (B)~\SI{1}{s}, (C)~\SI{3}{s}, and (D)~\SI{5}{s} expansion time in the disorder potential. Each image is averaged over six to ten realizations. Below and to the right of each panel, the corresponding one-dimensional integrated density profile $n_{1\rm d}(z,t)$ and $n_{1\rm d}(y,t)$ are shown. The progressive broadening of $n_{1\rm d}(z,t)$ illustrates the expansion dynamics. In contrast, because of the residual weak transverse confinement, the profiles along the $y$~axis show very little change. The dashed lines corresponds to a Gaussian fit of the transverse profiles,  the fitted rms size evolving from 9 to~\SI{10}{\micro \meter} (see Fig.~\ref{fig:transverseProfile}).}
		\label{fig:twoDexpansion} 
	\end{figure}

\begin{figure} 
			\centering
			\includegraphics[scale=0.36]{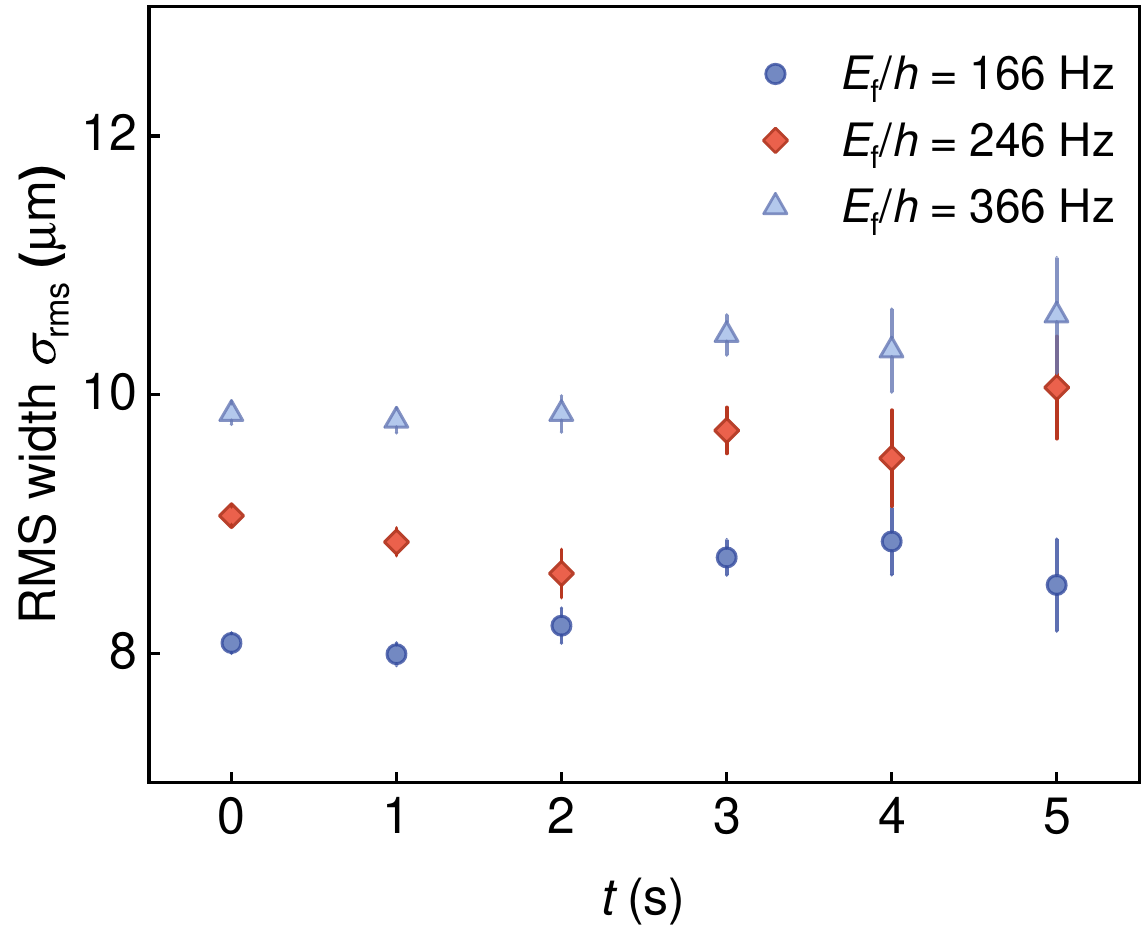} 
			\caption{\textbf{Transverse rms size along the $y$ direction.}~Time evolution of the transverse rms size $\sigma_{\mathrm{rms}}$ for three loading energies: $\Ef = 166$, $246$, and $326~\mathrm{Hz}$. Across the full energy range the transverse size shows only a weak increase from $t=0$ to \SI{5}{\second} of expansion in the disorder. Note that $\sigma_{\mathrm{rms}}$ remains always much larger that the estimated scattering mean free path $\ell \sim \SI{1.5}{\micro \meter}$, indicating that the transport dynamics remains three-dimensional (see Sec.~\ref{sec:confinement}). }
			\label{fig:transverseProfile} 
		\end{figure}

Figure~\ref{fig:twoDexpansion} displays representative 2D images obtained at $E_{\rm f}=\SI{246}{Hz}$ for $\VR/h$=\SI{416}{Hz} after $t=$\SI{0}{s}, $t=$\SI{1}{s}, \SI{3}{s}, and \SI{5}{s} of expansion within the disordered potential. As in the main text,  the 2D data are normalized for the same number of atoms for a proper comparison. The atomic cloud exhibits a clear expansion along the $z$ direction, while it is very limited on the transverse direction size because of the residual weak transverse confinement (see section~\ref{Sec:Levitation}). More precisely, the rms transverse size varies from about 9 to \SI{10}{\micro \meter} between 1 and \SI{5}{\second}. As shown in Fig.~\ref{fig:transverseProfile}, such very weak transverse evolution is observed from the low energy $E_{\rm f}=\SI{166}{\hertz}$ -- localised regime -- to the high energy $E_{\rm f}=\SI{366}{\hertz}$ -- diffusive regime. 

Note that the transverse profiles exhibit a slight asymmetry due to radiation pressure exerted on the atoms along the $y$ axis by the resonant beam used for fluorescence imaging (see Methods of the main text).

\subsubsection{Three-dimensional transport dynamics \label{sec:threeDdynamics}} 

At the numerically predicted mobility edge $E_c=240$ Hz, which sets the typical energy scale for the wave-packet spreading, we estimate the scattering mean free time to be $\tau\simeq 1\,$ms \cite{richard2019TauS}, and the corresponding scattering mean free path $\ell\simeq \sqrt{2E_c/m}\tau\simeq$ \SI{1.5}{\micro \meter}. This value is much smaller than the transverse confinement size $\sim$\SI{10}{\micro \meter}, indicating that the geometry of the system is not quasi-one-dimensional but genuinely three-dimensional. This conclusion is further supported by the excellent agreement between our observations and the numerically predicted position of the mobility edge. A detailed investigation of the role of the confinement is left for future work.





\subsection{Comparison with the self-consistent theory \label{Sec:SCT} }

The self-consistent theory of localization gives the following theoretical expression for the 1D density profile
\begin{align}
    n_\text{1d}(z,t) = \int^\infty_{-\infty} dz^\prime
     \int^\infty_0&  dE\, \mathcal{D}(E;\Ef) \\
    &\times P_{E^\prime}(z,z^\prime,t)n_\text{1d}(z^\prime,t=0) \;,  \nonumber 
    \end{align}
where $\mathcal{D}(E;\Ef)$ is the energy distribution of the atoms in the disorder for a given transfer energy $\Ef$, defined in Eq.~(\ref{eq:supEDistrib}), $P_{E}$ is the disorder-average propagator and $n(z^\prime,t=0)$ is the initial 1D density profile. Within this framework, the propagator is given by
\begin{align}
    & P_{E}(z,z',t) = \int_{-\infty}^\infty \frac{d\omega}{2\pi}\int_{-\infty}^\infty \frac{dq}{2\pi}\frac{e^{iq(z-z')-i\omega t}}{-i\omega+D(\omega)q^2} \nonumber \\
    &\!=\! \int_{-\infty}^\infty \frac{d\omega}{2\pi} \frac{e^{-i\omega t}}{\sqrt{-4 i\omega D(\omega)}}\exp\Big[\!-\!|z-z'|\sqrt{\frac{-i\omega}{D(\omega)}}\Big],
\end{align}
which involves a generalised diffusion coefficient $D(\omega)$ renormalised by quantum interference. This coefficient satisfies
\begin{align}
\label{eq:SCTL_D}
    \frac{1}{D(\omega)} &= &\frac{1}{D_B}
    +\frac{1}{\pi\rho\hbar D_B}
    \int^{Q_\text{UV}} \frac{d^3\boldsymbol{Q}}{(2\pi)^3}
    \frac{1}{-i\omega+D(\omega)\boldsymbol{Q}^2},  \nonumber \;
\end{align}
where $D_B$ is the classical (Boltzmann) diffusion coefficient, $\rho$ is the 3D density of state per unit volume and $Q_\text{UV}$ is a momentum cutoff used to regularize the theory at short spatial scales.
The solution of this equation can then be expressed as
\begin{equation}
\label{eq:X(omega)}
    D(\omega) = \left(\beta
    \frac{-\alpha(E)+[\sqrt{\alpha(E)^3-i\omega}+\sqrt{-i\omega}]^{2/3}}{[\sqrt{\alpha(E)^3-i\omega}+\sqrt{-i\omega}]^{1/3}}\right)^2,
\end{equation}
where $\beta=(8\pi^2\rho\hbar)^{-1/3}$ and $\alpha(E)=\alpha_0(1-E/E_c)=[Q_\text{UV}/(2\pi^3\rho D_B\hbar)-1]/(3\beta^2)$. In this formulation, the three physical quantities $\rho$, $D_B$ and $Q_\text{UV}$ are recast in terms of the three composite parameters $\alpha_0$, $\beta$ and $E/E_c$. The experimentally measured mobility edge $E_c/h = 237$ Hz is used as input. 

The parameters $\alpha_0$ and $\beta$ are determined by a global fit on the experimental peak densities, performing a simultaneous optimization over multiple energies and expansion times.
The optimal values are found by minimizing a global least-squares cost function defined as
\begin{equation}
    \chi^2(\alpha_0, \beta) = \sum_E \sum_t \left[n_{0,\text{exp}}(E,t) - n_{0,\text{SCT}}(E,t)\right]^2,
\end{equation}
where $n_{0,\text{SCT}}$ is the peak density predicted by the self-consistent theory for a given pair $(\alpha_0,\beta)$. The minimization is performed by a two-dimensional grid search over $(\alpha_0,\beta)$ and the best-fit parameters are identified as the global minimum of the resulting $\chi^2$ map: $\alpha = 3.17$ and $\beta = 6.06 \times 10^{-6}$ (SI units).


The initial density profiles $n(z^\prime,t=0)$ were determined by fitting the experimental density distributions right after the rf transfer from the disorder-free state $|1\rangle$ to the disorder sensitive state $|2\rangle$. To account for the slight asymmetry present in the data, the profiles were fitted using an asymmetric pseudo-Voigt function (see Ref.~\cite{C8AN00710A}). Examples of these fits are illustrated in Fig.~\ref{fig:InitialZprofile} at three loading energies.

\end{document}